\documentclass[12pt,preprint]{aastex631}

\usepackage{booktabs}
\usepackage{rotating}

\begin{document}
\title{Continuum And Line Emission From Accretion Shocks at T Tauri Stars I. Correlations With Shock Parameters}

\author{John Kwan}

\affil{Dept. of Astronomy, University of Massachusetts, Amherst,
MA 01003, jykwan@umass.edu}

\email{jykwan@umass.edu}

\begin{abstract}

Fourteen models are calculated
with the shock velocity ranging from 200 to 330 km s$^{-1}$ and
pre-shock hydrogen nucleon density ranging from
$2.5\times 10^{12}$ to $4\times 10^{13}$ cm$^{-3}$. Among them the summed
emergent flux of all spectral lines 
accounts for about 0.1-0.3 
of the total veiling flux. The
hydrogen Balmer continuum accounts for 0.17-0.1, while a
nearly constant fraction close to 0.5 comes from emission produced by the stellar
atmosphere. The main results derived from the veiling continuum energy
distributions are two strong correlations: 1) the Balmer jump (BJ) increases as $F_K$, the shock kinetic energy flux, decreases; 2) at a fixed fraction of surface coverage by accretion shocks $r_{\lambda}$, the ratio
of veiling to photospheric continuum flux at wavelength $\lambda$, decreases as
$F_K$ decreases. Using the BJ - $F_K$ and
$r_{4500\AA}$ - $F_K$ relations, the observed excess continua of 10 T Tauri stars are modelled. For BP Tau and 3 Orion stars our accretion luminosities are higher than published values by a factor of a few. For the 6 Chamaeleon I stars our observed accretion luminosities are
about 27 - 78\% higher than corresponding published values.  Comparison of model results on the HeI $\lambda$5876\AA$~$flux with observed data indicates that, while those stars with dominant $\lambda$5876\AA$~$narrow components can be readily accounted for by the calculated models, those with much stronger broad components cannot, and suggests that for the latter objects the bulk of their excess continua at 5876\AA$~$do not originate from accretion shocks.

\end{abstract}

\keywords{accretion, accretion disks -- line: formation -- radiative transfer -- shock waves -- stars: pre-main-sequence}

\section{Introduction}

The phenomenon of accretion onto a T Tauri star via infall from a
circumstellar disk along magnetic field lines is now well established
(Hartmann, Herczeg \& Calvet 2016). The key observation pointing to presence
of an accretion shock is the continuum excess over the photospheric
continuum that is readily detected at UV wavelengths through broadband
photometry and at optical wavelengths from the reduction of absorption
depths of photospheric lines in comparison with non-accreting weak
T Tauri stars (Basri \& Batalha 1990). Another observation signalling
infall is the presence of red absorptions spanning velocity widths of
$\ge$100 km s$^{-1}$ and extending to $\sim$300 km s$^{-1}$ in
certain spectral lines. A third observation that also suggests an 
accretion shock is the presence of prominent narrow components in 
some spectral lines, notably HeI $\lambda$5876\AA$~$ (Beristain,
Edwards \& Kwan 2001, hereafter BEK01)
and CIV $\lambda$1549\AA$~$ (Ardila et al. 2013).
Magnetic field strengths of several KG inferred from the wavelength
separation between right and left circularly polarized components of
the HeI $\lambda$5876\AA$~$ narrow profile in several T Tauri stars 
affirm the formation of the narrow component in close proximity to
the star (e.g. Johns-Krull et al. 2013). Theoretical calculation of line
formation in an accretion shock is still needed, however, to fully
verify the origin of the narrow components.

Detailed theoretical modelling of an accretion shock at a T Tauri star was
first presented by Calvet \& Gullbring (1998). They calculated the 
radiation spectrum produced by the post-shock gas, using a cooling
rate as a function of temperature generated from the work of Raymond \&
Smith (1977). Half of this radiation spectrum propagates towards the 
pre-shock zone and, through photoionization, is reprocessed into a
continuum spectrum, half of which emerges and forms part of the overall
veiling continuum. The other half of this continuum spectrum, together
with the remaining half of the above radiation spectrum generated in the
post-shock zone, propagates towards the star. Calvet \& Gullbring (1998)
calculated the emergent continuum spectrum produced by the stellar
atmosphere upon absorption of this incident radiation as well as the
flux emergent from the stellar interior below. This emergent continuum 
spectrum forms the remaining part of the veiling continuum. Their model
results with different shock parameters had been applied by them and
others (e.g. Ingleby et al. 2013, 2014) to determine accretion luminosities,
mass accretion rates, and surface covering factors of the accretion shocks
for a large number of T Tauri stars. Dodin (2018) had also presented
an accretion model. He calculated the evolution of the elemental
ionization structure and the cooling associated with collisional excitation
of energy levels of individual ions to obtain a self-consistent development
of the temperature and ionic fractions of an individual element as the
gas cools. He also determined the angular distribution of line emission
and his results focus on the line profiles produced in the post-shock and 
pre-shock zones, particularly for HeII $\lambda \lambda$1640, 4686\AA$~$
and HeI $\lambda$5876\AA.

In this work an accretion shock is also modelled. Besides continuum emission,
spectral line fluxes generated in both the post-shock and pre-shock zones
are calculated. The main motivation is finding out to what extent 
observed narrow and broad components of H, He, and metal lines originate
in accretion shocks. In particular, how well can the model account for
i) the relative prominence and range of equivalent widths among the
observed narrow components of HeI $\lambda \lambda$5876, 6678, 4471, 4713\AA,
HeII $\lambda$4686\AA$~$ (BEK01),
HeI $\lambda$10830\AA, 
P$\gamma$ (Edwards, Fischer, Hillenbrand \& Kwan 2006, hereafter EFHK06), H$\gamma$,
H$\delta$, CaII $\lambda \lambda$8662, 8498, 3970\AA$~$(Alencar \& Basri 2000);
ii) the observed fluxes and profiles among the UV lines HeII $\lambda$1640\AA,
CIV $\lambda$1549\AA, SiIV $\lambda$1397\AA, and NV $\lambda$1240\AA$~$
(Ardila et al. 2013). It is possible that the combined relations of line flux and continuum emission to model parameters will not only test the robustness of the accretion shock model but may also reveal new information on the observed excess continuum. Also, an understanding of the line emission from the post-shock and pre-shock zones may help unravel the contributions to observed line profiles from other kinematic regions,
as the diversities of profiles among observed optical and UV 
lines clearly point to two or more kinematic regions producing them.

Our theoretical model is described in this paper. Results on the veiling
continuum are also presented here,
together with comparisons between model and
observed veiling continua. HeI $\lambda$5876\AA$~$ is a prominent optical line produced in the post-shock zone and a large data sample of its flux and profile is available (BEK01), so the dependence of its flux on shock parameters is also presented here, while results on other prominent UV and optical lines are deferred to  a sequel. In the next section the procedures of calculating post-shock
cooling, pre-shock photoionization, and radiative transfer of line and
continuum emission are detailed. Section 3 provides examples of the model
veiling continuum, shows the dependence of the total veiling flux as
well as its constituents, like the Balmer and free-free continuum fluxes,
on physical parameters characterizing an accretion shock, and demonstrates
how the results can be used to deduce the shock kinetic energy
flux and fraction of stellar surface covered by the accretion shocks. Limitations of our model and possible improvements are also noted.
Applications of these results to 10 T Tauri stars are made in Section 4,
and the theoretical accretion luminosities compared with observed values. Section 5 presents the model HeI $\lambda$5876\AA$~$flux and its comparison with observed data.
In the last section the main results are recapped, followed by a discussion 
on the implications brought forth by comparisons between model and observed results on the nature of the excess continuum.

\section{Model}

The model presented here differs from earlier ones, Calvet \&
Gullbring (1998), Dodin (2018), primarily by calculating, in
addition to continuum flux, line fluxes emergent from an
accretion shock. Many lines are optically thick and a 
treatment of their radiative transfer is needed. Hydrogen and
helium produce prominent optical lines in the post-shock
cooling zone and their radiative recombinations to all the
energy levels included in the atomic models constitute part
of the emergent continuum, so both their line and continuum
emission are tracked. A second difference lies in the
demarcation of the post-shock cooling zone from the heated
atmosphere. This demarcation is not obvious in Calvet \& Gullbring (1998) as the radiation incident on their heated atmosphere consists of both ionizing photons generated at high temperatures and non-ionizing photons generated at lower ones. Robinson \& Espaillat (2019) updated the Calvet \& Gullbring (1998) accretion shock model by, instead of adopting a cooling function based on local thermal equilibrium, directly incorporating radiative cooling in solving the fluid equations for the temperature and ionization structures in the post-shock region. However, the region is assumed to be optically thin, so the associated ionizing and non-ionizing photons propagating towards the star also deposit all their energies in the heated photospheric region. Here the post-shock cooling calculation ends
when the ionized hydrogen fraction falls below 0.002, so all the ionizing photons propagating towards the star from both pre-shock and post-shock zones are absorbed, and only photons with energies less than 13.6 eV are incident on the heated atmosphere. This
ensures a careful accounting of the Balmer, Paschen, and Bracket continuum emission, as all ionizing photons are tracked and accounted for in the calculation. A third difference comes from the determination of the emergent spectrum from the heated atmosphere. Here the treatment is crude and the emergent spectrum is approximated by a black-body spectrum with total flux equal to the total incident flux. The results of the 14 models calculated also do not include the effect of the emergent flux from the stellar interior. For 3 models with low shock kinetic energy flux the calculations are repeated with the emergent black-body flux being the sum of $\sigma T_\ast^4$ and the incident flux. Calvet \& Gullbring (1998) calculate the emergent spectrum more accurately by modelling the radiative transfer of incident and stellar flux via the use of Rosseland mean opacities. Like Dodin (2018), the hydrodynamic equations are solved here together with equations governing the flux of elemental ions to determine the ionization and thermal structures in the post-shock region. An iterative procedure is followed here to include the effect of photons emitted from the pre-shock photoionization zone towards the post-shock zone. At the end of the process the total flux in emergent continua and spectral lines is compared with the input kinetic energy flux as a check on the accuracy of the model calculation. Below the general features of the model are described. Details of the calculation of post-shock cooling, pre-shock photoionization, and treatment of line and continuum radiative transfer are provided in following sub-sections.

The radiation produced in the theoretical model is tracked in detail
from 0.5 to 500 eV. From 0.5 to 5 eV and from 50 to 500 eV there are
500 energy bins each with bin width $\Delta h\nu = h\nu~ln(10)/500$.
In between, from 5 to 50 eV, 2500 bins with bin width $\Delta h\nu =
h\nu~ln(10)/2500$ are used, in order to accommodate the large number of
spectral lines that fall within that energy range. The energy flux of
photons with $h\nu \leq 0.5$ eV are summed together as a single entry
in bin 1. Similarly bin 3501 records the summed energy flux of photons
with $h\nu \geq 500$ eV.

There are two parameters that characterize each accretion shock model
calculation. They are the pre-shock velocity $u$, and pre-shock density,
signified here by $n_H$, the hydrogen nucleon density. Other elements 
included in the calculation are He, C, N, O, Ne, Si, Fe, Mg, and Ca with
assumed solar abundances (Lodders 2020). The pre-shock mass density is
then given by $\rho_o = \mu n_H m_H$ with $\mu = 1.355$. The values of $u$,
$n_H$, as well as $F_K = {1\over 2}\rho_o u^3$, an associated useful parameter
that represents the pre-shock kinetic energy flux, are listed in
Table 1 for each of the 14 models calculated. If the accretion shocks
cover one percent of the surface of a $2 R_\odot$ star, the listed
parameters indicate mass acccretion rates that range from
$5.2\times 10^{-9}$ to $1.2\times 10^{-7}$ M$_\odot$ yr$^{-1}$.

Plane-parallel geometry is assumed and the radiative flux in the forward
and backward directions are updated in each step of the calculation.
The post-shock cooling and pre-shock photoionization calculations end
when the ionized hydrogen fraction falls below 0.002 and 0.05 respectively.
By that time the temperature has fallen to $\leq$7000 K in either zone.
It turns out that the resulting hydrogen nucleon column density of
the post-shock cooling zone ranges from $2.8\times 10^{20}$ cm$^{-2}$
in model 10 to $2.4\times 10^{21}$ cm$^{-2}$ in model 1, while the column
length ranges from $4.9\times 10^4$ cm in model 5 to $2.2\times 10^7$
cm in model 11. The hydrogen nucleon column density of the pre-shock
photoionization zone is comparable, ranging from $1.6\times 10^{21}$ 
cm$^{-2}$ in model 10 to $1.5\times 10^{22}$ cm$^{-2}$ in model 1. The
column length, on the other hand, is much higher. Models with
$n_H = 2.5\times 10^{12}$ cm$^{-3}$ have the largest values,
spanning from $1.3 \times 10^9$ to $3.4\times 10^9$ cm as $u$ increases
from 240 t0 330 km s$^{-1}$. In comparison, the lateral dimension of
an individual accretion spot is $5\times 10^9$ cm, if it is 
assumed that 100 accretion spots together cover one percent 
of the surface of a 2$R_\odot$ star.
Thus, for the pre-shock photoionization calculation the approximation
of a plane-parallel geometry is poor for the lowest density models.
Nevertheless, it is made to make the calculation manageable.

To produce the final emergent line and continuum spectra, an iterative
procedure is adopted. For a given $u$ and $n_H$ the post-shock cooling
calculation is performed first with an
estimated ionization structure and temperature
of the pre-shock gas at the interface and no input radiative flux from the
pre-shock photoionization zone. 
At the end of the
calculation the forward radiative flux is assummed to be absorbed 
by the stellar atmosphere and
re-emitted as a thermal continuum. The latter is attenuated through
the post-shock cooling zone, using the calculated hydrogen continuum
opacites, to produce the emergent thermal spectrum at the shock interface.
This and the calculated backward radiative flux, which contains all the
ionizing flux beyond 13.6 eV, form the input radiation spectrum for
the subsequent pre-shock photoionization calculation. At the end of this
calculation the emergent line and continuum spectra are tabulated to
obtain for the first iteration the total veiling flux, as well as its 
constituents, like the hydrogen Balmer continuum flux, the total flux
of spectral lines, and the thermal continuum flux. The radiation spectrum
emitted towards the star from this iteration, and the calculated 
temperature and elemental ionization structure at the interface become
new inputs to the next iteration of post-shock cooling calculation.
The emergent thermal continuum spectrum at the interface is also an
input. Its flux, however, is tracked backwards because it flows against 
the advance of the post-shock cooling calculation. Thus, it is enhanced
rather than attenuated by the intervening hydrogen continuum opacities
between steps and its integrated total at the end of the calculation, in
comparison with the total forward radiative flux absorbed by
the stellar atmosphere, forms
one convergence criterion of the iterative process.

Among the constituents of the emergent veiling flux the thermal component
clearly increases the most from the first to second iteration because it
only has contribution from the post-shock cooling zone in the first
iteration. Typically it rises from $\sim$52\% of the final emergent
thermal flux to $\sim$93\% in the second iteration. The emergent Balmer
continuum flux also increases because the emission towards the star
from the pre-shock photoionization zone of HeII Lyman and Balmer 
continua, HeI Lyman continuum, HeII and HeI lines with $h\nu \ge$ 13.6 eV
will produce H and He photoionization in the post-shock cooling zone.
It typically rises from $\sim$ 80\% of the final emergent value in the
first iteration to $\sim$ 96\% in the next. To bring the iteration
process to a close, usually four or five iterations are needed for
the difference in each component of the emergent veiling flux to fall
below 0.5\% between successive iterations.

\subsection{Post-shock Cooling}

The steady-state hydrodynamic equations that govern the properties of 
the post-shock gas are:
\begin{equation}
	{d\over {dx}}(\rho v) = 0
\end{equation}
\begin{equation}
	{d\over {dx}}(P + \rho v^2) = 0
\end{equation}
\begin{equation}
	{d\over {dx}}[\rho v(\epsilon + {P\over {\rho}} + {1\over 2}v^2)] =
	-\Lambda~,
\end{equation}
where $x$ is the distance measured from the shock front, $v$, $\rho$, P,
and $\epsilon$ are the gas flow velocity, mass density,
pressure, and internal energy per unit mass 
respectively, and $\Lambda$ (ergs s$^{-1}$
cm$^{-3}$) is the rate of radiative energy loss. If the part of $\epsilon$
contributed by the translational degrees of freedom, which is ${3\over 2}
P/\rho$ for a monoatomic gas, is separated out from the rest,
denoted by $\epsilon_{iz}$, equation (3)
can be rewritten as
\begin{equation}
	{d\over {dx}}[\rho v({5\over 2}{P\over {\rho}} + {1\over 2}v^2)] =
	-\Lambda - {d\over {dx}}(\rho \epsilon_{iz} v)~.
\end{equation}

The immediate post-shock values of $\rho$, $v$, and $P$, which serve as
boundary conditions to the hydrodynamic equations, are related 
to the properties of the pre-shock gas through the above
equations by assuming  
no radiative loss or change in elemental ionization structure 
right at the shock front. If, in additional to the model parameters $n_H$
and $u$, $T_o$ denotes the temperature of the immediate pre-shock gas
and $\xi_e$ represents the ratio of the electron density to $n_H$, the
immediate post-shock velocity and temperature are:
\begin{equation}
	v_1 = {1\over 4}(1 + 5\eta) u
\end{equation}
\begin{equation}
	T_1= {{\mu m_H u^2}\over {16 k}} {{(3 - \eta)(1 + 5\eta)}\over
	{(1.085 + \xi_e)}}~,
\end{equation}
where $\eta = (1.085 + \xi_e) kT_o/(\mu m_H u^2)$. With typical values
of $T_o = 2\times 10^4$ K and $\xi_e = 1.1$, $\eta = 6.7\times 10^{-3}$
for $u$ = 200 km s$^{-1}$, so the highest value of the factor
$(1 + 5\eta)$ among the models is 1.033.

The elemental ionization structure changes as the flow advances from
the shock front, a consequence of collisional ionization primarily at
first and then photoionization and recombination subsequently. Its evolution
is followed at each step of the advance. Because the post-shock 
temperature can reach as high as $\sim 1.5\times 10^6$ K all the
ionization stages of H, He, C, N, O, and Ne are included. For Si and Fe
only the lowest 12 ionization stages, in addition to the neutral stage, are 
tracked. The elements Mg and Ca are not effective in cooling at
temperatures exceeding $3\times 10^4$ K. They are included for
the purpose of assessing
the strength of the MgII $\lambda$2798\AA, CaII H and K, and CaII infrared
triplet emission. For them no ionization stage beyond Mg III and Ca III is 
being considered.

To illustrate the calculation of the elemental ionization structure, the
procedure for carbon is selected. Letting $N_{CII}$ denote the number
density of C II, the change in the C II flux with distance is
\begin{equation}
	{d\over {dx}}(N_{CII} v) = - N_{CII}(N_e R_{CII} + N_e C_{CII} +
	\Gamma_{CII}) + N_{CI}(N_e C_{CI} + \Gamma_{CI}) + 
	N_{CIII} N_e R_{CIII}~,
\end{equation}
where $N_e$ is the electron density, $R_{CII}$ (cm$^3$ s$^{-1}$)
the recombination coefficient to C I, $C_{CII}$ (cm$^3$ s$^{-1}$)
the collisional ionization coefficient to C III, and $\Gamma_{CII}$ (s$^{-1}$)
the photoionization rate to C III. Equations for the flux of other
stages are analogous. The flux of carbon nucleons is conserved, i.e.,
\begin{equation}
	{d\over {dx}}(N_C v) = 0~,
\end{equation}
with $N_C$ being the sum of the number densities in all seven stages.
Upon being divided by $N_C v$, equation (7) specifies the change of the CII
fractional population with distance.

The term $d(\rho \epsilon_{iz} v)/dx$ in equation (4) is related to
equation (7). If $\epsilon_C$ represents the part of $\epsilon_{iz}$
contributed by carbon, the internal energy per unit volume due to carbon is
\begin{equation}
	\rho \epsilon_C = \sum_{J=II}^{VII} N_{CJ} E_{CJ}~,
\end{equation}
where $E_{CII}$,
for example, denotes the internal energy associated with the C II stage,
in reference to $E_{CI}$ being set to 0. If $\chi_J$ is the ionization
potential from stage J to J+1, with J=I referring to the neutral stage,
then $E_{CII} = \chi_I$ and
\begin{equation}
        E_{CVII} = \sum_{J=I}^{VI} \chi_J~. 
\end{equation}
The contribution of carbon to the term $d(\rho \epsilon_{iz} v)/dx$ is
then given by
\begin{equation}
	{d\over {dx}}(\rho \epsilon_{C} v) = \sum_{J=II}^{VII} E_{CJ}
	{d\over {dx}} (N_{CJ} v)~,
\end{equation}
and contributions of other heavy elements can be similarly written down.

In equation (7) the recombination coefficient has both radiative recombination
and dielectronic recombination components. The latter process is important
for many ions when the temperature exceeds $\sim 2\times 10^4$ K. At 
the high densities being considered here, however, collisional ionization
of the captured electron during its cascade to the ground state is 
significant (Davidson 1975). Rather than setting up a procedure for
reduction of the dielectronic recombination coefficient, calculations
will be made for the two extreme cases, one in which dielectronic
recombination is completely suppressed and the other in which it is 
incorporated fully.

Line emission constitutes a major contributor to radiative energy loss.
The atoms/ions involved in our model calculation are H I, He I$\rightarrow$II,
C II$\rightarrow$IV, N III$\rightarrow$V, O I$\rightarrow$VI, 
Ne V$\rightarrow$VIII, Si IV, Si VIII$\rightarrow$X, Fe II, Fe III,
Fe VIII$\rightarrow$X, Mg II, and Ca II. They are chosen to cover the
important cooling lines at temperatures ranging from $\sim 1.5\times 10^6$ 
to $\sim$6000 K. There are over 600 lines included among the heavy elements.
Most of them have photon energies exceeding 13.6 eV and are ultimately 
absorbed within the post-shock cooling and pre-shock photoionization
zones, thereby transferring their energies to continuum and line emission
at lower energies.

To calculate the line emission rates, the population in the different
energy levels of an atom/ion are assumed to be populated in statistical
equilibrium, expecting that collisional interactions between levels proceed
more rapidly than the rate at which the atom/ion density changes via
ionization or recombination. The hydrogen atom is selected to illustrate
the procedure. Assuming complete $l$-mixing and $k$ energy levels, the
population in level $i\geq2$, denoted by $N_i$ (cm$^{-3}$) is related to
the other level population and $N_{HII}$, the proton density, by equating
its rates of population loss and gain, as given by
\begin{eqnarray}
	N_i (N_e \sum_{j\neq i} C_{ij} + \sum_{j=1}^{i-1} A_{ij}\beta_{ij} +
	N_e C_{i\infty} + \Gamma_i) &=& \sum_{j\neq i} N_j N_e C_{ji} +
	\sum_{j=i+1}^k N_j A_{ji}\beta_{ji}\nonumber\\
	& &+ N_{HII} N_e (R_i - R_i^{\prime} + N_e R_i^{3b})~.
\end{eqnarray}
The analogous equation for the population in level 1 or the ground 
state is not adopted but replaced instead by the requirement that the
level population sum up to the neutral hydrogen density, i.e.
\begin{equation}
	\sum_{i=1}^k N_i = N_{HI}~.
\end{equation}
The fractional population among the levels can then be obtained once
$N_{HII}$ and $N_{HI} = N_{H} - N_{HII}$ are determined at a particular
step. 

In the above equations $C_{ij}$ (cm$^3$ s$^{-1}$) is the collisional 
coefficient from level $i$ to $j$, 
$A_{ij}$ (s$^{-1}$) the spontaneous emission rate to the lower level 
$j$ and $\beta_{ij}$ the corresponding escape probability, 
$C_{i\infty}$ (cm$^3$ s$^{-1}$) the collisional ionization
coefficient, $\Gamma_i$ (s$^{-1}$) the photoionization rate, $R_i$ (cm$^3$
s$^{-1}$) the radiative recombination coefficient to level $i$, and
$R_i^{3b}$ (cm$^6$ s$^{-1}$) the three-body recombination coefficient.
The use of the local escape probability method in place of the full radiative
transfer of spectral line photons will be described in a later
sub-section.

Twenty energy levels are employed for the hydrogen atom. At the model
$n_H$ values considered, collisional ionization and its reverse process,
three-body recombination, are increasingly important for highly excited
levels. The population of levels $i\geq 10$ are very close to values
in thermal equilibrium with respect to the proton density as a result. 
Only photoionizations from levels $i\leq 4$ are included, so $\Gamma_i = 0$
for $i\geq 5$. Ionizing photons originate from either the pre-shock
photoionization zone or the up-stream post-shock cooling region. They
are part of the line and continuum flux propagating in the direction 
aligned with the forward advance of the calculation. The step by step
nature of our calculation makes it difficult to include photons that
are emitted later downstream and propagating backwards. There is little
error caused by this omission for line photons that can ionize H I, He I,
or He II from the ground state, as they are all produced when those
continuum opacities are very much less than unity. For continuum radiation 
this omission is partly redressed by the following procedure, as 
illustrated with the Lyman continuum. Half of those photons propagate
backwards and their attenuation in passing through the up-stream
post-shock gas is assumed to be effected right at the same emission spot.
Thus, if $\tau_1$ denotes the Lyman continuum opacity measured from the
shock front to the calculation point at the threshold frequency $\nu_1$,
the on-the-spot absorption coefficient (cm$^3$ s$^{-1}$) is
\begin{equation}
	R_1^{\prime} = {1\over 2} (1 - e^{-\tau_1
	{(\nu_1 / \bar{\nu_1})}^3}) R_1~,
\end{equation}
where $h\bar{\nu_1}$  is the average energy of a Lyman continuum
photon obtained from ratioing the total energy to total number of 
photons emitted. This on-the-spot approximation is fair for the Lyman
continuum as its opacity, once nearing unity, rises rapidly to a very high
value, but crude for the Balmer, Paschen, and Brackett continua whose
opacities through the whole post-shock cooling zone are less than unity.
For the remaining hydrogen continua no attenuation is considered, so
in equation (12) $R_i^{\prime}$ is set to 0 for $i \geq 5$.

The calculation of the He I or He II energy level population is analogous.
Nineteen levels are employed for He I and ten for He II. For both of
them there is no photoionization from excited levels, and no attenuation 
of continua emitted from radiative recombinations to excited levels,
so $\Gamma_i$ and $R_i^{\prime}$ are 0 for $i \geq 2$. Corresponding
equations for calculating the fractional population among energy levels
of a heavy ion are even simpler. Except for Ca II, the coefficients
$C_{i\infty}$, $\Gamma_i$, $R_i$, $R_i^{\prime}$, and $R_i^{3b}$ in
equation (12) are
all set to 0, so only collisional and radiative interactions among the
energy levels are included. For Ca II, modelled by a three-level system,
only $R_i^{\prime}$ and $R_i^{3b}$ are ignored. More details on
calculating the emission of the Ca II lines, as well as the Fe II and 
Fe III UV multiplets, will be provided in paper II.

Collisional ionization and photoionization from excited states of hydrogen
not only affect the level population and thereby line emission but also
the hydrogen
ionization structure. Corresponding to equation (7) for a heavy ion, the
change in H I flux with distance is
\begin{equation}
	{d\over dx} (N_{HI} v) = - \sum_{i=1}^k N_i (N_e C_{i\infty} + 
	\Gamma_i) + N_{HII} N_e \sum_{i=1}^k (R_i -
	R_i^{\prime} + N_e R_i^{3b})~.  
\end{equation}
As noted above, $\Gamma_i$ and $R_i^{\prime}$ are 0 for $i\geq 5$.
In the same way equations for the changes in He I and He II fluxes
can be written down. For them $\Gamma_i$ and $R_i^{\prime}$ are 0
for $i \geq 2$.

Once the elemental ionization structures and fractional population 
among energy levels are obtained $\Lambda$, the rate of radiative energy
loss, is readily determined. Both line and continuum emission contribute
to $\Lambda$ positively, while continuum absorption via photoionization
contributes negatively. As an example, the contribution from hydrogen
emission is
\begin{equation}
	\Lambda_H = \sum_{i=2}^k \sum_{j=1}^{i-1} N_i A_{ij} \beta_{ij}
	h\nu_{ij} + \sum_{i=1}^k N_{HII} N_e (R_i - R_i^{\prime})
	h\bar{\nu_i} - \sum_{i=1}^4 N_i \int_{\nu_i}^{\infty}
	\sigma_i (\nu) f_{\nu} d\nu~,
\end{equation}
where $h\nu_{ij}$ is the photon energy of the $i\rightarrow j$ transition,
$\sigma_{i} (\nu)$ the level $i$ photoionization cross-section and 
$f_{\nu}$ (ergs s$^{-1}$ cm$^{-2}$ Hz$^{-1}$) the radiative flux density
propagating forwards. The latter includes all the line and continuum 
radiation incident from the pre-shock photoionization zone and the
forward-propagating line and continuum radiation generated in the
up-stream post-shock cooling region. It also includes the thermal
continuum which produces ionizations of hydrogen from levels $i = 2-4$.
As mentioned earlier this particular component, propagating in the
opposite direction, is tracked backwards.

Analogous equations apply for the contribution to $\Lambda$ from each
atom/ion of the other elements. They are simpler because  
$R_i^{\prime}$ = 0 for each heavy ion while only
$R_1^{\prime}$ $\neq$ 0 for He I and He II. Then, except for Ca II, 
photoionization takes place only from the ground state. Another 
simplification is made for every heavy ion except CaII in that 
continuum photons produced from recombinations are taken to each have an
energy of $h\nu_1$. The great majority of these photons are more
energetic than 13.6 eV, so they are ultimately absorbed within the 
post-shock cooling or pre-shock photoionization zone and the simplification
will not introduce a discrepancy in the overall energy budget.

There is also contribution to $\Lambda$ from free-free emission.
It is
\begin{equation}
	\Lambda_{ff} = 1.426 \times 10^{-27} T^{0.5} N_e
	(N_{HII} G_{HII} + N_{HeII} G_{HeII} + 4 N_{HeIII} G_{HeIII})~,
\end{equation}
where $T$ (K) is the temperature, $N_{HeIII}$ (cm$^{-3}$) the He III 
number density, $G_{HeIII}$ the He III
Gaunt factor, etc. A polynomial with four parameters is used to fit 
the theoretical Gaunt factors of Sutherland (1998) as a function of
log ($\chi / kT$) with $\chi$ being the ionization potential which
is 13.6, 24.58, and 54.4 eV for $G_{HII}$, $G_{HeII}$, and $G_{HeIII}$
respectively.

With $\Lambda$ and $d(\rho \epsilon_{iz} v)/dx$ procured
from gathering up their respective constituents at a particular
distance $x$ , the right hand side of equation (4) is determined.
Employing the conservation of mass flux (eq. [1]) and of momentum
flux (eq. [2]) to relate $\rho$ and $P$ to $v$ on the left hand side,
the value of $v$ at $x + dx$ is found. This procedure is repeated
once in that after $v$ and the other gas properties are obtained at
$x + dx$ the right hand side of equation (4) is evaluated there and
its average with the previous value at $x$ is used to produce the
revised and adopted values of $v$, $P$, etc. at $x + dx$. Concurrently
line and continuum opacities are updated, forward continuum flux 
densities attenuated, and backward thermal flux densities enhanced.
Then, with the adopted temperature and elemental population of
energy levels, line and continuum emission at $x + dx$ are calculated
to augment the line and continuum fluxes propagating forwards and
backwards.

A simple check of the overall accuracy of the post-shock cooling
calculation is provided by the conservation of total energy flux,
as equation (3) indicates. When the latter is integrated over
distance the change in energy flux carried by the post-shock gas 
between the shock front and the end point of calculation is 
simply equal to the total rate of radiative energy loss suffered by
the gas in traversing the distance, which, in turn, is just the
energy flux carried by the line and continuum radiation emerging from
both ends. The calculation also involves incident radiation from the
pre-shock zone at the beginning point and from the stellar atmosphere
at the other end. They can also be readily incorporated into the balance
of energy flux. Thus the sum of the energy fluxes carried by the
post-shock gas at the beginning point and the incident radiation from
both ends equals the sum of the energy fluxes carried by the gas at
the end point and all the radiation, including incident components
propagating through, emergent from both ends. This equality is realized
to within 0.5\% among the computed models.

\subsection{Pre-shock Photoionization}

Heating of the pre-shock gas through photoionization produces a
temperature no higher than $\sim 2.5\times 10^4$ K among the computed
models. This considerably simplifies the pre-shock photoionization
calculation. Thus atoms/ions whose spectral lines provide cooling are 
fewer here. They are H I, He I$\rightarrow$ II, C II$\rightarrow$ IV,
N III$\rightarrow$ V, O III$\rightarrow$ VI, Si IV, Fe II, Fe III,
Mg II, and Ca II, and, for each of the heavy ion, fewer energy levels
need to be included. Also, for Ne, Si, Fe, Mg, and Ca, only the ionization
stages Ne I$\rightarrow$ III, Si I$\rightarrow$ IX, Fe I$\rightarrow$ V,
Mg I$\rightarrow$ IX, Ca I$\rightarrow$ IV are involved.

The difference in pressure of the pre-shock gas between 
the shock interface and the end point of the photoionization 
calculation is small compared with the kinetic momentum flux $\rho_o u^2$,
so the calculation is carried out at constant density.
Ionization equilibrium at each point is imposed. Derivation of the
fractional population of each element in its different ionization stages
follows that outlined earlier for the post-shock calculation, with the
additional simplification of setting the left hand sides of the relevant
equations (cf. eqs [7], [15]) to 0. Determination of the population
among the energy levels within a particular ionization stage is also
identical to that described in the post-shock calculation. Finally thermal
equilibrium is assumed at each point, and the temperature is found by
balancing radiative cooling from line and continuum emission against
photoionization heating to bring the radiative loss term $\Lambda$ to 0.
This requires an interative process involving adjustments of $T$ from an
initial guess and, to a lesser degree, the electron fraction $\xi_e$ 
since elemental ionization structures, as well as incremental line
and continuum opacities and attenuation of the forward-propagating
radiation, have to be updated each time. When $T$ changes from its
previous value by $\leq$ 0.5\% between successive iterations, their average is
adopted as the correct value. The same criterion applies concurrently
to $\xi_e$. Line and continuum emission are then evaluated to augment
the radiative fluxes propagating forwards and backwards.

At each step of the calculation the radiative energy flux propagating
forward and that emergent
from the other end are summed and compared with that originally incident
to monitor the accuracy of the computation procedure.
At the end of the calculation this sum deviates from the incident energy
flux by $\leq 1.7$\% among the computed models.

In both the post-shock cooling and pre-shock photoionization calculations
atomic data parameters, such as collisional ionization and photoionization
cross-sections, are needed. Their collection is described in the
Appendix.

\subsection{Radiative Transfer}

The simple step by step advance of our calculation procedure to determine
gas properties and emission can only treat radiative interactions of
line and continuum photons crudely and approximately. For spectral lines
the local escape
probability method employed by Kwan \& Krolik (1981) in their photoionization
model of the emission lines in active galactic nuclei is followed.
The local line emission rate (cm$^{-3}$ s$^{-1}$) from an upper level $i$
to a lower level $j$ is given by $n_i A_{ij} \beta_{ij}$ with $\beta_{ij}$
being the probability of photon escape from the calculation zone. The
latter clearly depends on the opacity of the spectral line between the
local point and each end of the zone, i.e.,
\begin{equation}
	\beta_{ij} = {1\over 2} \beta (\tau_{ij}) + {1\over 2} 
	\beta (\tau_{ij}^{\prime})~,
\end{equation}
with $\tau_{ij}$ being the line opacity measured from the beginning end.
As the calculation advances $\tau_{ij}$ is directly procured. For
$\tau_{ij}^{\prime}$, given by $\tilde{\tau_{ij}} - \tau_{ij}$, the 
total line opacity $\tilde{\tau_{ij}}$ is first set to an arbitrarily
high value in the first run of either post-shock cooling or pre-shock
photoionization calculation and subsequently revised.

The adopted expression for $\beta (\tau)$ is based on calculations that
determine the probability of a photon escaping from the mid-plane of a
uniform slab of total optical depth $2\tau$ at line center (Avery \& House
1968, Adams 1972). It is
\begin{equation}
	\beta (\tau) = {1\over {\tau \pi^{0.5} \{ 1.2 +
	0.5 [ln (\tau +2.3)]^{0.5} / (1 + 10^{-4}\tau)\} }}~,
\end{equation}
for $\tau > 0.5$. It attempts to join smoothly the two expressions 
obtained when frequency redistribution in the Doppler core and in 
the damping wings dominates respectively, taking $10^4$ as the opacity
demarcating the two regimes. When $\tau \leq 0.5$ $\beta (\tau)$ is 
simply set equal to 1.

In the post-shock cooling zone it turns out that the great majority of
spectral lines of heavy elements providing cooling at temperatures 
$\geq 3 \times 10^4$ K are optically thin. Of those with photon energies
$\geq$ 13.6 eV the most notable exception is the very strong line
Ne VIII $\lambda$ 774 \AA$~$ which has the highest $\tilde{\tau_{ij}}$,
but of a value ranging from only $\sim$ 2 to $\sim$ 8 among the computed
models. These photons also cause ionizations. Those escaping towards the
star are absorbed further downstream at temperatures $<3 \times 10^4$ K
when the He and H continuum opacities rapidly rise. Those escaping
towards the pre-shock zone are not attenuated as the relevant continuum
opacities are $\ll$ 1. However, as part of the incident radiation
for the pre-shock photoionization calculation their energies are
ultimately converted into emergent non-ionizing ($h\nu < 13.6$ eV) photons.
Thus, in the overall transformation of kinetic energy into emergent photons
through the intermediate step of collisional excitation of energy levels
of heavy elements, the crude treatment of the radiative transfer of line
photons more energetic than 13.6 eV introduces little error. Among 
important non-ionizing cooling lines of C, N, O, Ne, and Si the very
strong line OVI $\lambda 1034$ \AA$~$ has an opacity $\sim 4.5$ 
among the computed models, while the others have opacities $\leq 2$
except for CIII $\lambda$ 977 \AA, CII $\lambda$ 1335 \AA, and
CII $\lambda$ 1036 \AA $~$ which are very optically thick. The strong
FeII UV multiplets, MgII $\lambda 2798$\AA, as well as most H and He
lines, which grow in strength at temperatures $< 3\times 10^4$ K, are
also very optically thick. The emergent flux of such a line with a high
$\tilde{\tau_{ij}}$ will be less accurate, but it does account for the
bulk of its photons that escape because it is generated primarily at
small $\tau_{ij}$  where the temperature is higher than that at 
$\tau_{ij} \sim \tilde{\tau_{ij}}$.

In the pre-shock photoionization zone the comparatively lower temperatures
produce little ionizing photons through collisional excitation. Those
emitted belong primarily to He I and He II permitted transitions from
upper states, populated as a result of recombination and cascade, down to
the ground state. These transitions are very optically thick and the
photons escape mostly towards the star and photoionize H and He in the
post-shock zone. Important non-ionizing spectral lines of H, He, and
heavy elements that provide cooling are all very optically thick. Their
emergent fluxes are uncertain because of both our crude treatment of 
their radiative transfer and inaccuracies in the $\tilde{\tau_{ij}}$  
values. More specifics on individual lines will be given in Paper II.

Radiative transfer of continuum photons is modelled with greater care. 
In the post-shock zone photons are emitted over a very broad temperature
range. Continuum photons emitted backwards will become incident
radiation on the pre-shock zone and need to be tracked accurately.
For the H Lyman, Balmer, Paschen, Brackett, He II Lyman and
Balmer continua the energy distribution of each of them is calculated
at each step, taking into account attenuation by the relevant intervening
continuum opacity, to augment its emergent energy distribution from
the shock front. At the end of the post-shock cooling calculation the
total emergent flux in each continuum is computed and compared with
that obtained using the simplified method of determining intervening
attenuation via use of the average energy of a continuum photon in the
on-the-spot approximation (cf. eq. [14]). This comparison is very close,
with the H Lyman continuum flux having the largest mis-match of
$\sim$ 4\%. The emergent energy distribution of each continuum is then
renormalized so its total emergent flux agrees with the value obtained in
the on-the-spot approximation. For the free-free continuum its energy
distribution is also calculated at each step. No attenuation of its 
emission towards the pre-shock zone is made, however, as the intervening
H, He I, and He II Lyman continuum opacities are tiny when free-free
emission at their respective energy thresholds are significant. For the
He Lyman continuum its energy distribution is not calculated. Rather, its
flux emitted at each step is taken to be all in photons of $h\nu = 24.58$
eV. This simplification is made because the emission is produced 
primarily at temperatures of kT $<$ 2 eV, so its energy distribution has 
only a narrow spread beyond the threshold limit.

Radiative transfer of continuum photons emitted in the forward direction
is simpler in the post-shock zone. For each of the H Lyman, Balmer,
Paschen, Brackett, He II Lyman and Balmer continua only the forward total
energy flux and total photon number flux are separately updated at
each step for the purpose of determining the average photon energy for
calculating appropriate attenuation as the calculation advances. The 
free-free continuum, however, has a broad energy distribution and its flux
in each energy bin from 0.5 to 500 eV is updated and attentuated
accordingly. The thermal continuum flux in each energy bin is also 
back-tracked and enhanced at each step.

In the pre-shock photoionization calculation continuum photons emitted
in the forward direction will contribute to the eventual emergent spectral
energy distribution for photons of energies $\leq 13.6$ eV. Accordingly, for
each of the H Balmer, Paschen, Brackett, and free-free continuum emission
its forward flux in each energy bin is tracked at each step as it is
being attenuated and augmented. For each of the H Lyman, He I Lyman,
He II Lyman and Balmer continuum emission the simplified treatment of
radiative transfer is adopted via updating at each step only the 
forward total energy and total photon number fluxes. The procedure is 
further simplified for the He I and He II continua by taking the energy
flux emitted to be all in photons at their respective threshold energies.
For the continua emitted backwards towards the shock front the 
determination of their emergent fluxes is analogous to that described
earlier in the post-shock cooling calculation for the corresponding backward 
emergent fluxes. There is the additional simplification in that only for
the H Lyman and Balmer continua are their energy distributions fully
calculated. For the H Paschen and Brackett continua only the backward
total energy flux and total photon number flux are separately updated
at each step, while for the He I Lyman, He II Lyman and Balmer continua
their total backward fluxes are taken to be all in photons at their 
respective threshold energies. For the free-free continuum also, only its
total backward emergent flux is updated as no attenuation of it is
included in propagating through either the pre-shock zone or subsequently
the post-shock zone.

Line and continuum radiation emergent from the shock front and incident
on the pre-shock zone are attenuated in propagating through as a result
of photoionizing He II and He I from their ground states, H I from
energy levels $i=1-4$, heavy atoms/ions from their ground states, and
Ca II from its two excited levels in addition to the ground state. At 
the end of the pre-shock photoionization calculation the emergent spectral
lines, H Balmer, Paschen, Brackett, and free-free continua that originate from
the post-shock zone are isolated to distinguish them from the corresponding
contribution generated by the pre-shock zone.

\section{Continuum Emission And Shock Parameters}

Post-shock cooling and pre-shock photoionization produce both line and
continuum emission. Photons emerging have energies less than 13.6 eV
owing to the high HI Lyman continuum opacity within the accreting 
column. Emergent spectral line fluxes will be presented and discussed in
Paper II, but their summed total, $F_{slns}$ (ergs s$^{-1}$ cm$^{-2}$), is
provided here to show their combined contribution to the total emergent
flux. Continuum emission include contributions from bremsstrahlung encounters 
between electron and individually $H^+$, $He^+$, and $He^{++}$,  
radiative recombinations of $H^+$ and $e$ to energy levels $n\geq 2$, of
$He^+$ and $e$ to $n\geq 2$, and of $He^{++}$ and $e$ to $n\geq 3$. For the
free-free continuum and each of the HI Balmer, Paschen, and Brackett continuum
the spectral energy distribution is computed in addition to the emergent
flux, denoted by $F_{ff}$, $F_{Bal}$, $F_{Pas}$, and $F_{Bra}$ respectively.
For the remaining continua from radiative recombinations only their summed
emergent flux is produced, given by $F_{rtrec}$. As specified in $\S$ 2, 
the total incident flux
absorbed by the stellar atmosphere is assumed to be re-emitted as
a thermal black-body. Its spectral energy distribution, after propagating
through the two zones, is also determined, together with its integrated
flux $F_{thrm}$. The emergent free-free, Balmer, Paschen, Brackett, and
thermal continua summed together forms the veiling continuum produced by
the accretion shock in our model. Its spectral energy distribution
$f_{\lambda}^V$ (ergs s$^{-1}$ cm$^{-2}$ \AA$^{-1}$) spanning in wavelength from
912 to 24800 \AA, the fluxes $F_{slns}$, $F_{ff}$, $F_{Bal}$, 
$F_{Pas}$, $F_{Bra}$, $F_{rtrec}$, $F_{thrm}$, and the list of individual
spectral line fluxes constitute our primary results.

For each of the 14 models listed in Table 1 two separate calculations are
performed, one with complete suppression of dielectronic recombination
(case a) and the other with no suppression at all (case b). When their results
have obvious differences both will be shown in side by side panels.

The total veiling flux $F_V$, given by the sum of $F_{slns}$, $F_{thrm}$,
$F_{Bal}$, $F_{Pas}$, $F_{Bra}$,
$F_{rtrec}$, and $F_{ff}$, can be directly compared with
$F_K$ since post-shock cooling, pre-shock photoionization, and re-processing
of absorbed radiation by the stellar atmosphere effectively convert the
input kinetic energy flux to emergent radiative flux. Our model $F_V$ value 
is slightly off from the corresponding $F_K$ value, and their difference 
varies largely with $u$. As the latter runs through 330, 300, 270, 240, and
200 km s$^{-1}$, the ratio $F_V/F_K$ equals 1.02, 1.03, 1.04, 1.05, and
1.08 respectively in case a and 0.97, 0.99, 1.01, 1.03, and 1.05 
respectively in case b. It needs to be noted that $F_V$ is inherently an
overestimate of $F_K$ because in the pre-shock photoionization zone whereas the
velocity and density are taken to be constant in our model the velocity is
actually lower at the shock front where photoionization commences
than at the location where photoionization effectively ceases. This is
because at the commencement point 
the gas is fully ionized and the temperature about $2\times 10^4$ K
while at the cessation point the gas is neutral and the temperature
considerably lower, so the associated pressure gradient decelerates the gas
as it moves through the photoionization zone towards the shock front. The
kinetic energy flux to be compared with $F_V$ should then employ the velocity  
at the cessation point. When this is taken into account the appropriate $F_K$
is about 0.5, 0.57, 0.66, 0.8, and 1.0\% higher for $u=$ 330, 300, 270, 240,
and 200 km s$^{-1}$ respectively. The differences between $F_V$ and $F_K$
range from -3 to 7\% among the models. They indicate largely the inaccuracies
of our numerical procedure.

\subsection{Veiling Continuum and its Components}

Figure 1 illustrates the emergent energy distributions obtained using 
case a results for the 3 $n_H$ values of $u$ = 300 km s$^{-1}$. Besides the
energy distribution of the veiling continuum (solid line), those of its 
constituents, namely, the thermal continuum (dotted line), the H continuum
comprising the Balmer, Paschen, and Brackett continua (dashed line), and the
free-free continuum (dot-dashed line) are also shown. The most conspicuous
feature in the veiling continuum is the Balmer jump, which increases in
magnitude as $n_H$, and $F_K$ accordingly, decreases. The Paschen and Brackett
jumps are also noticeable. All of them are, of course, much more prominent
in the plot of the H continuum by itself. The thermal and free-free continua,
on the other hand, show Balmer, Paschen, and Brackett drops as they are
absorbed at the ionization thresholds.

It can be gauged that the thermal continuum is the major constituent of the
veiling continuum, followed by the H continuum. This is more clearly 
demonstrated in Figure 2, where the fractional contribution of an individual
radiative component to the total emergent flux, given by $F/F_V$ with $F$
standing for $F_{thrm}$, $F_{slns}$, $F_{Bal}$, $F_{ff}$, or $F_{rtrec}$, is
plotted against $F_K$. Instead of the H continuum the Balmer continuum is
singled out here because it is more easily extracted from observed data. The
Paschen and Brackett continua will behave similarly, as
$F_{Pas}/F_{Bal}$ and $F_{Bra}/F_{Bal}$ are, within $\pm$ 4\% among the models,
constant at 0.333 and 0.138 respectively. All 14 model results are exhibited,
with each of the 5 lines plotted for an individual radiative component showing
the variation of $F/F_K$ with $n_H$ at constant $u$, whose value can be
deduced from each line's right end point.

Over the range of $F_K$ explored the thermal continuum accounts for close to
a constant fraction of 0.5 of the emergent flux. It turns out that the hydrogen
continua have opacities less than unity and most of the dominant emergent lines
have opacities less than a few, so it is not surprising that roughly half of
the radiative flux generated in the post-shock cooling and pre-shock 
photoionization zones propagate towards the star. Thus in case a the Balmer
opacity has the highest value in model 1, equalling 0.266 and 0.45 in the
post-shock cooling and pre-shock photoionization zone respectively, and it
scales with $u$ and $n_H$ roughly as $u^a n_H^b$ with $a 
\simeq$4.4, $b \simeq$0.8 in the post-shock cooling zone, and
$a \simeq$2.9, $b \simeq$0.15 in the pre-shock photoionization
zone. The $F_{thrm}$ presented represents the emergent value, after its
attenuation through the two zones, so the thermal continuum flux produced by
the stellar atmosphere is actually slightly higher.

The next significant components of $F_V$ are the spectral lines collectively
and the Balmer continuum. The prominent emergent spectral lines originate
from the post-shock cooling zone, including OVI $\lambda$1034\AA,  
CIV $\lambda$1549\AA, CIII $\lambda$977\AA, and FeII UV multiplets as a  
whole. Their effectiveness as coolants diminishes when the temperature exceeds
$\sim 0.7\times 10^6$ K. As $u$ increases from 200 to 330 km s$^{-1}$, the
peak post-shock temperature rises from $0.57\times 10^6$ to $1.5\times 10^6$
K. Consequently the fractional contribution of spectral lines to the 
veiling flux, $F_{slns}/F_V$, falls as $u$ increases. Conversely 
$F_{Bal}/F_V$, as well as $F_{ff}/F_V$ and $F_{rtrec}/F_V$, rises with 
increasing $u$, since the effective cooling lines at temperatures $\geq 
0.7\times 10^6$ K have photon energies above 13.6 eV and they ultimately
produce ionizations of H and He, leading subsequently to free-bound and
free-free emission.

At a fixed $u$ $F_{slns}/F_V$ also falls as $n_H$ increases, but the decline
is gentler. Using case a to illustrate, a factor of 4.3 increase in $F_V$, via
$u$ increasing from 200 to 330 km s$^{-1}$, produces an increase in $F_{slns}$
by a factor of $\sim 2.0$. On the other hand, a factor of 4 increase in $F_V$
via increasing $n_H$ produces an increase in $F_{slns}$ by a factor of 
$\sim 3.2$. This weaker dependence of $F_{slns}/F_V$ on $n_H$ is largely 
because in the post-shock cooling zone   
OVI $\lambda$1034\AA, NV $\lambda$1240\AA, and CIV $\lambda$1549\AA$~$ have
opacities that increase little with increasing $n_H$.
Over an increase in $n_H$ by a factor of 16, their combined flux
increases almost proportionally, by a factor of $\sim 13.5$. On the other
hand the FeII UV multiplets and hydrogen lines are optically thick. 
Their combined flux increases by a factor of only $\sim 5.5$. Summing
all the spectral line fluxes together, $F_{slns}$ increases by a factor of
$\sim 10$.

Case a and case b clearly have their biggest contrast in the degree of
contribution from spectral lines. The full incorporation of dielectronic
recombination in case b causes more rapid $ion + e$ recombinations among
heavy elements. The lower OVI, NV, CIV abundances as a result reduce the
efficacies of OVI $\lambda$1034\AA, NV $\lambda$1240\AA, and 
CIV $\lambda$1549\AA$~$ emission as coolants and produce a smaller
$F_{slns}/F_V$. The contrast between the two cases is stronger when $u$ is 
lower, reflecting the greater effectiveness of those line emission in
cooling as the temperature falls from $1.5\times 10^6$ K.

As mentioned earlier, $F_{Pas}/F_{Bal}$ and $F_{Bra}/F_{Bal}$ are nearly
constants, so the plots of $F_{Pas}/F_V$ and $F_{Bra}/F_V$ with respect
to variations in $u$ and $n_H$ will be almost identical to those of
$F_{Bal}/F_V$. Together the three H continua account for 14.4 to 23.2 \%
of $F_V$ in case a and 19.2 to 24.6 \% in case b within the covered
parameter space. Over a factor of 41.6 in the span of
$F_K$ their fractional contribution varies little comparatively.

The magnitudes of $F_{Bal}$, $F_{Pas}$, $F_{Bra}$, $F_{rtrec}$, and
$F_{ff}$ depend on the strengths of their respective emission rate
coefficients. Their responses to variations in $u$ and $n_H$, however,
are all similar, as each of them depends on density squared, and their
combined flux is tied to $F_{thrm}$ and $F_{slns}$ through the total 
sum being $F_V$. Slight differences among the $F_{Bal}/F_V$, $F_{rtrec}/
F_V$, and $F_{ff}/F_V$ plots seen in Figure 2 arise primarily from the
different dependences of the respective continuum emission on temperature.

The free-free continuum flux accounts for $\leq 0.12$ of the total 
veiling flux. Its spectral energy distribution, as seen from Figure 1, is 
also relatively flat at optical wavelengths. It will be difficult to
discern from observational data its presence as a constituent of the veiling
continuum.

Fluxes from H, He, and He$^+$ recombination continua add up to $F_{rtrec}$.
Their respective fractional contributions are roughly 0.62$\pm$0.03,
0.33$\pm$0.04, and 0.05$\pm$0.02 among the 14 models in both cases a and
b. The hydrogen continua contribute at $\lambda$ longward of $\sim 2\times
10^4$\AA. Their presence will not be readily isolated from the stronger
underlying stellar continuum.

Parts of the He$^+$ continua have thresholds identical to the H Balmer,
Paschen, and Brackett edges. Incorporating their 
spectral energy distributions, however, will add no more than 0.6\% to
the H continuum plots in Figure 1. Among the included He continua the one
produced from recombinations to energy level $2p~^{3}P^o$ is the strongest.
It has a threshold at $\lambda=3436$\AA, close to the H Balmer edge, and
a strength $\sim$6\% of the Balmer continuum.

It should be noted that, whereas the emergent spectral line flux originates
almost exclusively from the post-shock cooling zone, $F_{Bal}$, $F_{Pas}$,
$F_{Bra}$, $F_{rtrec}$, and $F_{ff}$ are generated in both zones. Thus each 
of the noted continuum flux has two components whose strengths and 
dependences on $u$ and $n_H$ are different because the two zones have
very different temperature distributions and ionization structures. 
Figure 3 shows the fractional share contributed by the post-shock
cooling zone to $F_{slns}$, $F_{ff}$, and $F_{Bal}$. Over 90\% of the 
emergent spectral line flux originates there. This is because in the
pre-shock photoionization zone the important cooling lines have high 
opacities, causing their photons to escape preferentially in the 
direction towards the star. Those with $h\nu \geq$ 13.6 eV, like
HeII $\lambda$304\AA, and HeI $\lambda$584\AA, will cause photoionizations
in the post-shock cooling zone while the rest will be absorbed by the 
stellar atmosphere, ultimately contributing to the emergent thermal
continuum. Their combined flux is $\sim$ 5-8 times the flux
of spectral lines emergent from the pre-shock photoionization zone.

The post-shock cooling zone is also the bigger contributor to $F_{Bal}$,
$F_{Pas}$, $F_{Bra}$, and $F_{ff}$. The ionizations of He and H there
are effected first by collisions with electrons and subsequently 
through photoionizations. The flux of spectral line photons producing
photoionizations is about the same for both zones. The temperature in the
pre-shock photoionization zone, on the other hand, is $\leq 2.5\times 10^4$ K,
much less effective in producing collisional ionizations. The much higher
range of temperature present in the post-shock cooling zone also favors
free-free emission over free-bound recombination, and produces a higher
fraction of $F_{ff}$ than $F_{Bal}$.

\subsection{Diagnostics of $F_K$ and $\delta$}

From Figure 1 it is seen that the most conspicuous feature of the veiling
continuum is the Balmer jump, and that its magnitude appears to increase
as $n_H$, and correspondingly $F_K$, decreases. The Balmer jump (BJ) is defined here as the ratio of the flux in energy bin just above the Balmer ionization threshold to that in the energy bin just below. This apparent trend is 
explored fully in Figure 4 which plots BJ as a function
of $F_K$ for both cases a and b. The individual points with the same
model parameter of $n_H$ are linked together to aid visual inspection.
Because each model result is tied to its model $F_V$ and, as noted 
earlier, the discrepancy between $F_V$ and $F_K$ varies among the models, 
for uniformity the value of $F_K$ plotted is set to the value of $F_V$
in this and subsequent plots.

Figure 4 clearly shows that BJ depends primarily on $F_K$. Thus models
5 and 6 have $u$ equal to 200 and 330 km s$^{-1}$ respectively but
close values of BJ($F_K$), being 1.9($3.91\times 10^{11}$) and
1.74($4.16\times 10^{11}$ ergs s$^{-1}$ cm$^{-2}$) respectively in case a.
The same situation is true for models 10 and 11. This direct dependence
of BJ on $F_K$ can actually be anticipated from Figure 2. There it is seen
that both $F_{thrm}$ and $F_{Bal}$ have close to a linear dependence on $F_K$
while the dispersions caused by variations in $u$ are comparatively small.
Hence BJ is a good diagnostic of $F_K$ but not $u$ or $n_H$ individually.

The rise of BJ as $F_K$ falls is readily understood from the $f_{\lambda}$
plots of the thermal and H continua in Figure 1. The H continuum 
$f_{\lambda}$ shows that at the Balmer edge the jump up from the
Paschen continuum is big and slightly higher as $n_H$ decreases. When 
$F_{Bal}$ falls, its flux density at the Balmer edge drops roughly 
proportionally, as seen from Figure 1. The corresponding flux density of
$F_{thrm}$, on the other hand, responds much more sensitively. To
illustrate, with $n_H$ decreasing from $4\times 10^{13}$ to 
$2.5\times 10^{12}$ cm$^{-3}$ in Figure 1, $F_{thrm}$ falls by a 
factor of 16, its black-body temperature drops by a factor of 2, from 
$10^4$ to $5.1\times 10^3$ K, but its $f_{\lambda}$ at the Balmer edge
drops by a factor of 45. This strong sensitivity of the thermal
continuum flux density at 3648\AA$~$ to its black-body temperature 
over the explored parameter space is largely responsible for
the higher BJ in the veiling continuum 
as $n_H$, and correspondingly $F_K$, decreases. A second contributing 
factor is the lower Balmer continuum opacity through the two zones,
which reduces the Balmer drop in the thermal continuum, as $u$ and
$n_H$ decrease.

Comparing the veiling and thermal continuum plots in Figure 1, it is seen
that shortward of 3648\AA$~$ the thermal continuum contributes only a
fraction of the veiling continuum, as the Balmer continuum is a
significant contributor. Longward of 3648\AA$~$, however, the thermal
continuum provides the bulk of the veiling continuum. Since $F_{thrm}$
depends primarily on $F_K$ the veiling continuum flux density $f_\lambda^V$ at  
a wavelength $\lambda>3648$\AA$~$ 
will also be a useful signpost of $F_K$.
Its utilization, however, involves knowledge of the surface area covered
by the accretion shocks. Taking the latter to be $\delta 4\pi R_\ast^2$,
with $\delta$ being the area covering parameter,
the ratio of the veiling continuum flux density to underlying stellar
continuum flux density $f_\lambda^\ast$
is $r_{\lambda}=\delta f_{\lambda}^V /f_{\lambda}^\ast$. Approximating
$f_\lambda^\ast$ by a black-body of temperature $T_\ast$,
Figure 4 plots $r_\lambda$ at $\lambda$=4500\AA, denoted
by $r_B$, versus $F_K$ for $\delta$=0.01 and $T_\ast$ = 3200, 3600, 4000,
and 4300 K. It is evident that, besides BJ, $r_B$ is another good diagnostic
of $F_K$ that observational data can provide.

Making use of the BJ versus $F_K$ and $r_B$ versus $F_K$ plots together,
both $F_K$ and $\delta$ can be deduced. Once the underlying stellar 
continuum is extracted from the observed continuum and the veiling
continuum uncovered the BJ in the veiling continuum and the observed 
$r_B$ value are obtained. The former points directly to $F_K$ which, in
conjunction with the effective temperature of the underlying stella 
continuum, points to the model $r_B$ value for $\delta$=0.01. The ratio
of the observed $r_B$ to model $r_B$ times 0.01 then yields $\delta$ for
the observed star.

Figure 4 shows the BJ - $F_K$ and $r_B$ - $F_K$ relations for both cases
a and b. They are very similar. In deducing accretion luminosities and
$\delta$s for a selection of 10 stars in the next section the case a
relations will be applied.

The $r_B$ - $F_K$ relation is chosen together with the BJ - $F_K$ relation
to deduce $\delta$ because, as seen from Figure 1, the veiling
continuum is strong at 4500\AA$~$ and may be more readily separated from
the underlying stellar continuum there. In principle other $r_\lambda$ -
$F_K$ relations can be used. Figure 5 shows such relations for $\lambda$=5700,
$\delta$ = 0.01 and considering 
$T_\ast $ values of 3600 and 4000 K. The ratio of two
$r_\lambda$s is independent of $\delta$ and is therefore a direct diagnostic
of $F_K$ once the effective temperature of the underlying stellar continuum
is determined. This is illustrated in Figure 5, using $r_B/r_R$ with $r_R$
denoting $r_\lambda$ at $\lambda$=5700\AA. There is an anti-correlation
between BJ and $r_B/r_R$ in that a higher BJ entails a lower $r_B/r_R$.
This can provide a self-consistency check on the model calculations.

\subsection{Selected Model Results}

In Figures 2-5 model results have been connected together to aid visual 
inspection of relationships. They are actually discrete entities of the
14 models. Several important ones are listed here in Table 2 for ready
reference.

\subsection{Model Refinements}

The BJ - $F_K$ relation indicates that BJ increases as $F_K$ decreases.
It reaches a value of $\sim$ 6 at the lowest value of $F_K$ investigated. Observed BJs from data samples of Gullbring et al. (2000),
Ingleby et al. (2014), and Manara et al. (2016), however, are less than $\sim$ 3.
The high model BJ values at low $F_K$s are likely caused in part by not taking account of the emergent flux from the stellar interior at the accretion 
spot. When the flux of the veiling thermal component, 
$F_{thrm}$, is not much higher
than $\sigma T_{\ast}^4$, the stellar atmosphere will be heated comparably
by emergent flux from below and incident flux from the post-shock zone
above. A better approximation of the emergent thermal continuum at the 
accretion spot is then a single black-body component with a combined
flux of ($F_{thrm} + \sigma T_{\ast}^4$) rather than two components with
two black-body temperatures as tacitly assumed in the presented model
calculations. To explore the consequential effects on BJ and $r_B$,
models 9, 12, and 14 of case a are repeated with the
emergent stellar interior flux included in the calculation. Two $T_{\ast}$
values are considered. Cases 1 (C1) and $1^{\prime}$ (C$1^{\prime}$)
have $T_{\ast} = 4000$ K and cases 2 (C2) and $2^{\prime}$ (C$2^{\prime}$)
have $T_{\ast} = 3600$ K. C$1^{\prime}$ and C$2^{\prime}$ refer to new
calculations with $\sigma T_{\ast}^4$ included while C1 and C2 refer to
the original ones. The $F_{thrm}$ values of models 9, 12, and 14 
correspond to black bodies of temperature 6167, 5146, and 4340 K
respectively. When $\sigma T_{\ast}^4$ is added, with $T_{\ast} = 4000$ K,
the resulting thermal continua have temperatures of 6423, 5562, and 4971 K.
Emergent fluxes of other veiling constituents, namely $F_{Bal}$,
$F_{slns}$, etc, change little, by $\leq 1.0$ \%. The thermal continuum
flux density at the Balmer continuum threshold is very sensitive to its
temperature, however, so BJ is the most affected. Table 3 lists the
resulting BJ and also Paschen jump (PJ) in each of the different cases.
It is seen, as expected, that if the C$1^{\prime}$  and C$2^{\prime}$ BJs
are plotted in Figure 4 the rise of BJ with decreasing $F_K$ is slower,
more so for $T_{\ast} = 4000$ K. The variation of PJ with $F_K$ within
each individual case is itself weak, and so is its variation between
cases.

Besides BJ, $r_{\lambda}$ is another result that depends on the veiling
continuum level and hence also responds sensitively to the inclusion 
of the emergent stellar interior flux. Assuming a $\delta$
of 0.01 for the accretion shocks, $r_B$ and $r_R$ in the same 4 cases
are also listed in Table 3. As expected, $r_B$ is more strongly affected
than $r_R$, as the veiling continuum level at 4500\AA$~$ is more sensitive
to the thermal temperature than that at 5700\AA.

The above discussion brings out a shortcoming of our model, the crude
treatment of the response of the stellar atmosphere to the incident
radiation from the accretion shock. At the end of the post-shock cooling
calculation the spectrum of photons less energetic than 13.6 ev 
propagating towards the star is known. A stellar atmosphere subroutine
that calculates the emergent stellar spectrum can be modified to include
the above incident spectrum as additional input and produce accurately
the resulting emergent continuum energy distribution, particularly at
the Balmer ionization threshold. The two BJs obtained in Table 3
without and with $\sigma T_{\ast}^4$ included as emergent flux likely
constitute upper and lower bounds, as the assumption of plane-parallel
geometry and one-dimensional radiative transfer ignores the angular spread
in photon emission. In all likelihood the radiation from an accretion spot
propagating towards the star is incident upon an area larger than the
actual size of kinematic impact, so the temperature characteristic of 
the resulting emergent thermal continuum will lie between the two 
black-body estimates obtained without and with $\sigma T_{\ast}^4$ added
to $F_{thrm}$.

This
additional uncertainty in the determination of $F_K$ and $\delta$ can be ameliorated in future model applications.
The simple procedure used earlier in Section 4 can provide
initial estimates of $F_K$ and $\delta$. Thereupon the whole veiling
continuum  energy distribution $f_{\lambda}^V$ from a model or from
interpolation between models can be combined with the adopted stellar
continuum template to produce fits to the actual observed continuum
spectrum to refine those estimates and maybe even the extinction measure.
The model output includes the energy distribution of the thermal component,
the total incident flux on the stellar atmosphere, and the total Balmer,
Paschen, and Brackett opacities through the post-shock and pre-shock
zones. The emergent stellar interior flux, $\sigma T_{\ast}^4$, can then
be readily included, and the new $f_{\lambda}^V$ used to assess 
additional error margins of $F_K$ and $\delta$ inherent in our model
calculations.

\section{Comparison Of Model Continuum Results With Observations}

In this section the model BJ - $F_K$ and $r_B$ - $F_K$ relations are
applied to 10 T Tauri stars to determine for each of them the accretion
luminosity $L_{acc}$ and surface covering factor $\delta$ of the accretion
shocks. Gullbring et al. (2000) have assembled optical spectra from
Gullbring et al. (1998) and UV data from IUE Archive and procured the
observed spectral energy distribution of BP Tau from 1400 to 5400\AA,
making this star a prime candidate for theoretical modelling. The 
observations of Ingleby et al. (2014) extend in wavelength from 9500\AA$~$
shortwards to 3300\AA, and include 1800-2600\AA$~$ UV data
for a couple of objects. From their sample the stars SO 1036, SO 540,
and CVSO 58 are selected for the broad range covered by their BJs.
Manara et al. (2016) have observed a large sample of stars and presented
for each the reddening-corrected observed spectrum as well as the underlying
stellar continuum template and the excess or
veiling continuum energy distribution. From
their sample TW Cha, T3, and VW Cha have a broad range in their BJs and
the same effective stellar temperature of 4060 K, and are selected. Three
other stars, ESO H$\alpha$562, T3-B, 
and T40 also have a broad range in their BJs, but their
effective stellar temperatures range from 3415 to 3780 K, and they are
also included. In addition to these 10 stars whose veiling continua are
being modelled there are 14 stars with $r_\lambda$ values determined
for $\lambda$ between 4883 and 10830 \AA$~$ (Fischer, Edwards, Hillenbrand 
\& Kwan 2011). These $r_\lambda$ versus $\lambda$ plots are also compared
with the ones produced by accretion shock models.

Table 4 lists the important stellar properties of the 10 selected stars
collected from the above-mentioned references. They include stellar distance
$d$, radius $R_\ast$, and effective temperature $T_\ast$ of the underlying
stellar continuum. When the item is not directly available, it is 
determined by making use of other information present, such as stellar
spectral type and luminosity. The flux density ratio $r_B$ of the veiling 
continuum to underlying stellar continuum at 4500\AA$~$ is also estimated
from visual measurements off the pertinent plots published, and listed.

The presence of an accretion disk implies that the observed stellar solid angle $\Omega$ will depend on viewing angle and the size of the disk truncation gap. Letting $\Omega$ be given by $\epsilon\pi R_\ast^2/d^2$,
the viewing factor $\epsilon\leq 1$ can be estimated by comparing the 
observed flux density of the adopted underlying stellar continuum with
that expected from an unobscured one. Making the comparion at 6000\AA$~$
where approximation of the unobscured stellar continuum by a black-body
of temperature $T_\ast$ is suitable, this viewing factor
is calculated for BP Tau, SO 1036, SO 540, and
CVSO 58. For the six objects selected from Manara et al. (2016) the
comparison is made at 4600\AA$~$ because the plotted spectra do not
extend beyond 4630\AA. This estimate is denoted by $\epsilon_{bb}$ 
to indicate its approximation of the correct $\epsilon$ 
obtained with the adopted stellar continuum template. Its value for
each star is listed in Table 4.

The viewing factor is necessary to convert the model result $F_K$ to an
accretion luminosity for comparison with the observed value. The latter
has been determined by Manara et al. (2016) for each of their observed
stars via use of a multi-grid fitting procedure in conjunction with a slab 
model of the hydrogen continuum emission to estimate, in addition to
requisites like visual extinction and appropriate underlying stellar 
template, the total veiling flux from that observed between 3300 and
7150\AA$~$ and thence the observed accretion luminosity. The latter is 
denoted by $L_{acc}^{obs}$, and its value obtained by Manara et al. (2016)
is listed for each of their six stars selected.

The observed BJ in the veiling continuum of a star points to $F_K$ and
from the $r_B - F_K$ relation the area covering factor $\delta$ can be
found. The observed accretion luminosity from the theoretical model to be
compared with the listed value is then given by $L_{acc}^{obs} = F_K\delta
\epsilon(4\pi R_\ast^2)$. If the stellar surface obscured by the accretion
disk is similarly covered by accretion shocks, the accretion luminosity
for the entire star is then given by $L_{acc} = F_K\delta(4\pi R_\ast^2)$.
Gullbring et al. (2000) and Ingleby et al. (2014) report such values
using the theoretical models of Calvet \& Gullbring (1998) and those
values for the four stars selected are listed.

Approximation of the underlying stellar continuum by a black-body continuum
means that the values of $L_{acc}^{obs}$ and $L_{acc}$ produced by our
modelling of the observed veiling continuum are actually given by
$F_K\delta_{bb}\epsilon_{bb}(4\pi R_\ast^2)$ and 
$F_K\delta_{bb}(4\pi R_\ast^2)$ respectively. If, at the wavelength
chosen to determine the viewing factor and
at 4500\AA$~$, the black-body flux density is
$\alpha$ and $\beta$ times respectively that of the adopted stellar template, 
then the correct viewing factor is $\epsilon = \alpha\epsilon_{bb}$, and
the correct area covering factor is $\delta = \delta_{bb}/\beta$. Our
model values of $L_{acc}^{obs}$ and $L_{acc}$ will need to be multiplied
by $\alpha/\beta$ and $1/\beta$ respectively.

The correlation of BJ with $F_K$ is tight, particularly for $u$ between
240 and 330 km s$^{-1}$. An observed BJ will point directly to a $F_K$ via
interpolation between the model results. This procedure is not followed
here in view of the approximate estimates of $\epsilon$, BJ, and $r_B$
from published plots. Instead the model with its BJ closest to the observed
value is chosen to simulate the observed veiling continuum. The
corresponding model number, BJ, and $F_K$ being set equal to $F_V$, as
well as the derived values of $\delta_{bb}$, $L_{acc}^{obs}$, and $L_{acc}$,
are listed for each selected star in Table 4.

\subsection{BP Tau}

Gullbring et al. (2000) show in their Figure 1 the underlying stellar
continuum employed, the veiling continuum from the theoretical modelling
of Calvet \& Gullbring (1998), and their sum to compare with the
de-reddened observed spectral energy distribution from 1400 to
5400\AA. Also shown are their two constituents of the veiling continuum,
which are the emission from the heated photosphere below the shock and
emission from the pre-shock and attenuated post-shock regions. Following
their lead, Figure 6 plots a black-body spectrum of temperature 4000 K,
the veiling continuum produced in model 3 multiplied by a factor
$\delta_{bb} = 7.72\times 10^{-3}$, and their sum. To compare directly
with the corresponding plots in Figure 1 of Gullbring et al. (2000),
the vertical scale of Figure 6 needs to be multiplied by the factor
$\epsilon_{bb} R_\ast^2/d^2 = 4.27\times 10^{-20}$. Thus at 6000\AA$~$
the black-body flux density, the veiling continuum flux density, and
their sum will have observed values of $5.12\times 10^{-14}$,
$1.79\times 10^{-14}$, and $6.92\times 10^{-14}$ ergs s$^{-1}$ cm$^{-2}$
\AA$^{-1}$ respectively. Also plotted in Figure 6 are the two 
constituents of the veiling continuum, namely the thermal continuum, and
the sum of the H continuum and free-free continuum.

Comparing the black-body spectrum with the underlying stellar continuum 
template indicates that their difference begins to 
occur at $\sim$ 4000\AA$~$ and increases towards shorter wavelengths.
At 6000 and 4500\AA$~$ the corresponding values $\epsilon_{bb}$ and 
$\delta_{bb}$ are reasonable estimates of $\epsilon$ and $\delta$ then.
In overall shape from 1400 to 6000\AA$~$ the total continuum spectrum
shown in Figure 6 appears to reproduce that actually observed fairly
well. Gullbring et al. (1998) have presented the extracted excess
spectrum of BP Tau in finer detail from 3200 to 5300\AA$~$ in their
Figure 7. To compare with it, the bottom part of Figure 6 plots on the
same scale the model veiling continuum distribution after multiplication by
the factor $\epsilon_{bb} R_\ast^2/d^2$ to bring it into the observed
realm. Longwards of 3700\AA$~$, it reproduces the observed excess continuum
distribution well, but shortwards of the BJ, the model veiling flux
density is about 10\% higher.

Examining the model results in greater detail through the veiling 
continuum constituents, it is seen that the BJ in the model H continuum
is considerably higher than that present in the H continuum of Gullbring
et al. (2000) that comprises pre-shock and attenuated post-shock emission.
At the same time emission from the heated atmosphere has a Balmer drop in
our model but a BJ in theirs. These differences most likely arise from
the different demarcations of the post-shock cooling zone from the
heated atmosphere. In our model the post-shock cooling zone ends when
the hydrogen ionization fraction drops below 0.2\%. In the case of BP Tau
when this occurs the temperature is $6.75\times 10^3$ K, the hydrogen 
Lyman continuum opacity and nucleon column density as measured from the shock
front are $6\times 10^3$ and $1.1\times 10^{21}$ cm$^{-2}$ respectively.
This demarcation ensures that all the ionizing photons propagating
towards the star are absorbed and their heating effect and the ensuing
hydrogen recombination properly taken account of within our numerical
calculation procedure. If Calvet \& Gullbring (1998) end their post-shock
region earlier, a fraction of the ionizing photons will photoionize 
hydrogen in their heated atmosphere zone and the emission from there will 
contain Balmer continuum photons and exhibit a BJ.

As gauged from Figure 6 the model veiling continuum appears to match the
observed excess continuum about as well as that presented by Gullbring
et al. (2000). The derived values of $F_K$ and $L_{acc}$, however, are 
quite different. They are $9.23\times 10^{11}$ ergs s$^{-1}$ cm$^{-2}$ 
and 0.444 $L_\odot$ from our model versus $3.16\times10^{11}$ ergs s$^{-1}$
cm$^{-2}$ and 0.138 $L_\odot$ from theirs. The area covering factor is
similar though, being $7.72\times 10^{-3}$ and $7\times 10^{-3}$
respectively.

\subsection{SO 1036, SO 540, CVSO 58}

Ingleby et.al. (2014) have presented in their Figure 4 the de-reddened 
observed CTTS spectrum and adopted WTTS template for each of their eight
observed stars, and in their Figure 5 the corresponding extracted excess
continuum. From these two figures the quantities $r_B$ and $\epsilon_{bb}$
are estimated for each of the three selected stars. Then to simulate
the excess continuum, a theoretical model is chosen based on the magnitude 
of its BJ. The model number, BJ, $F_K$ = $F_V$ value,
and the derived $\delta_{bb}$, together with $r_B$ and $\epsilon_{bb}$,
are listed in Table 4. Two stars have stellar templates with
an effective temperature of 4000 K while the third, SO 1036, has
3700 K, so the $\epsilon_{bb}$ and $\delta_{bb}$ values may not
be far off. Figure 7 plots
for each star the predicted excess continuum distribution that will be
observed according to the parameters $R_\ast$, $d$, and $\epsilon_{bb}$
on the same scale as that used for the corresponding observed
spectrum. Also plotted is the excess continuum obtained by Ingleby et al. (2014), upon application of the Calvet \& Gullbring (1998) accretion shock models, that provides the best fit to the observed data. For SO 540 the actual observed excess continuum extends in wavelength from 8000 to 1800 \AA, but for SO 1036 or CVSO 58 it only extends to 3300 \AA. Overall, the comparison between model and observed spectra is fair.

Model 9 is used to simulate the excess continuum of SO 1036. The $F_K$ value of this model is low and taking account of the emergent stellar interior flux is relevant. The results illustrated in Table 3 indicates that for $T_\ast$ = 3600 K, which is close to SO 1036's, the new BJ is smaller than the original one by 9.4$\%$ and the new $r_B$ is higher than the original one by 13.6$\%$, so the corresponding $\delta_{bb}$ will be smaller by the same percentage than the original one listed in Table 4.

Ingleby et al. (2014) have also plotted in their Figure 3 the veiling
$r_{\lambda}$ as a function of wavelength for each of their observed
stars, employing two different WTTS as templates for unveiled absorption
lines. To compare with the data points of the three selected stars, Figure
8 plots on the same wavelength scale the $r_{\lambda}$ obtained in the
model using a black-body spectrum to approximate the underlying stellar
template, and the fit to the data points derived by Ingleby et al. (2014). Given the large scatter of the data points, being more than $\pm 25\%$ about the fit, the predicted dependence is in reasonable agreement with that observed.

The stars SO 1036, SO 540, and CVSO 58 are selected for the decreasing
order in BJ magnitude seen in their respective veiling continuum. 
According to the BJ - $F_K$ relation the derived $F_K$ is consequently
in ascending order. The area covering factor is arbitrary, so there is
no trend in the derived $\delta_{bb}$ and thence $L_{acc}$. Comparing 
these model results, listed in Table 4, with the corresponding values
of Ingleby et al. (2014) is somewhat hampered by their use of two
components, one of high $F_K$ and the other of $F_K$ 10-100 times lower.
For SO 1036 our $F_K$ (ergs s$^{-1}$ cm$^{-2}$) of 1.64 $\times$ 10$^{11}$
is roughly the geometric mean of their two components of 10$^{12}$
and $10^{10}$. Our $L_{acc}$ ($L_\odot$) of 0.53 is quite a bit higher
than their 0.1 value because our single component has an area covering
factor of 2.44 $\times$ 10$^{-2}$,
about half of that occupied by their $F_K$ = 10$^{10}$ component
but much higher than the factor of 2 $\times$ 10$^{-4}$ for their $F_K$ =
10$^{12}$ component. For SO 540 our $F_K$ of 3.14 $\times$ 10$^{11}$
is close to their high $F_K$ component of 3 $\times$ 10$^{11}$, but
has an area covering factor of 5.67 $\times$ 10$^{-3}$ that is $\sim$ 11
times higher. Our $L_{acc}$ of 0.112 happens to be close to their 0.1
value because their low $F_K$ component of 3 $\times$ 10$^{10}$ has an
area covering factor of 4 $\times$ 10$^{-2}$ and contributes over 85\%
of their $L_{acc}$. For CVSO 58 the comparison between the theoretical 
models is the closest. Their high $F_K$ component of 10$^{12}$,
with an area covering factor of 5 $\times$ 10$^{-3}$, provides the bulk 
of a $L_{acc}$ of 0.3, while our model has a $F_K$ of 1.66 $\times$
10$^{12}$, an area covering factor of 8.56 $\times$ 10$^{-3}$, and a
$L_{acc}$ of 0.5.

\subsection{Chamaeleon I Stars}

Manara et al. (2016) have observed and analyzed 34 Chamaeleon I young
stellar objects and presented in their Figure C.1 the photospheric
template, the veiling continuum from their slab model, and their sum 
to compare with the reddening-corrected observed spectrum for 32 of
them. From this group of 32 stars six are selected for modelling. Among
the remaining stars nine have $T_\ast \leq $ 3270 K and derivations of 
$\epsilon$ and $\delta$ using a black-body spectrum to approximate the
actual photospheric template will be quite inaccurate shortwards of
4600\AA. Another ten have either very weak veiling continua or very
small BJs to model reliably. For the six selected stars their pertinent
data parameters and model results are listed in Table 4.

Simulation of the veiling continuum of each selected star follows the 
same procedure as described earlier for BP Tau and the three Orion stars.
Figure 9 presents for each of TW Cha, T3, and VW Cha the predicted
veiling continuum observed in accordance with the values of $R_\ast$,
$d$, and $\epsilon_{bb}$. Following the convention of Manara et al.
(2016), it is in units of ergs s$^{-1}$ cm$^{-2}$ nm$^{-1}$ and plotted
over the same wavelength range from 330 to 470 nm. These three stars all
have a photospheric template with effective temperature of 4060 K, 
and are arranged from top to bottom in order of decreasing BJ. In the
same way Figure 10 presents the model veiling continua of ESO H$\alpha$562,
T3-B, and T40. Their photospheric effective temperatures are different,
being 3705, 3415, and 3780 K respectively.

A look at Figures 9 and 10 indicates that the overall shape of the veiling
continuum correlates with the BJ magnitude. Thus, at $\lambda <$ 364.8 nm
the slope of the veiling continuum towards shorter wavelengths change 
from rising to falling as BJ increases, while at $\lambda >$ 364.8 nm the
slope towards longer wavelengths change from falling to rising, though 
less noticeably. This correlation is actually more conspicuous in Figure 7
where the plotted wavelength range, from 1800 to 8000\AA$~$, for the
three Orion stars is more extensive. It is also evident from the veiling
continua illustrated in Figure 1 for the three models with $u$ = 300 km
s$^{-1}$ and different $n_H$ values. The comparison of model results with
a large number of observational data brings this correlation more into 
focus. It also sets up a strict test of the theoretical model.

Overall the theoretical veiling continua of the six selected stars compare
reasonably well with the corresponding observed excess continua shown in Figure C.1
of Manara et al. (2016) that are generated by their slab model through a
multi-grid fitting procedure involving also the visual extinction and
photospheric template. From Table 4 it is seen that the derived viewing 
factors cluster between 0.6 and 0.7 except for T3-B which has $\epsilon_{bb}$
= 0.42 and also the lowest $T_\ast$ of 3415 K among them. Most likely the
black-body spectrum overstimates the photospheric template at $\lambda \leq$
460 nm, and the derived $\epsilon_{bb}$ will be an underestimate. For the
same reason the derived $\delta_{bb}$ will be an overestimate. 
Interestingly, because $L_{acc}^{obs}$ is given by
$F_K\delta_{bb}\epsilon_{bb}(4\pi R_\ast^2)$ its derived value will be 
less affected, as $\delta_{bb}$ and $\epsilon_{bb}$ are evaluated at fairly
close wavelengths. The derived $L_{acc}$, on the other hand, will be 
affected just as $\delta_{bb}$.

Comparing the model $L_{acc}^{obs}$ with the corresponding value of
Manara et al. (2016) the former is higher in each case, by a factor 
ranging from 1.26 for VW Cha to 1.79 for ESO H$\alpha$562. One obvious
cause for this result is the inclusion of spectral line fluxes in the
total veiling flux in our model. An additional cause may come from He
and He$^+$ recombination and free-free continua being part of our
veiling continuum, but they may not be a part of the continuum 
generated by the slab model of Manara et al. (2016). The two causes
together can lead to our $L_{acc}^{obs}$ being higher by a factor of
$\sim$ 1.3. Taking this factor into consideration, the closeness between
the observed and model values is fairly remarkable.

Among the remaining stars in the sample of Manara et al. (2016) several,
like CT Cha A, have BJs smaller than the smallest BJ obtained in the
14 models. An additional model with $u$ = 300 km$^{-1}$ and $n_H$ =
10$^{14}$ cm$^{-3}$ is run, and it has a BJ of 1.21 and $F_V$ of
3.11$\times10^{12}$ ergs s$^{-1}$ cm$^{-2}$, so the BJ - F$_K$ 
relation will extend.

In producing model veiling continua to compare with observed data, case a
results have been used. If, for each of the ten selected stars, the same 
model is used but with case b results the derived $\delta_{bb}$ is $\sim$
3-5\% higher than, while $L_{acc}$ is within $\pm$ 1\% of the corresponding
case a values.

\subsection{$r_\lambda$ versus $\lambda$ Plots of 14 Stars}

Fischer, Edwards, Hillenbrand \& Kwan (2011) have presented in their
Figure 1 $r_\lambda$ versus $\lambda$ plots of 14 stars in their effort to
determine the excess continuum emission from 0.48 to 2.4 $\mu m$ of
their sample of 16 classical T Tauri stars. The value of $r_\lambda$ depends
on the underlying stellar continuum energy distribution which is approximated
by a black-body spectrum in our model, and two $T_\ast$ values, 3600 and
4000 K, are chosen to illustrate. It also depends on the surface covering
factor $\delta$ of the accretion shocks and, to eliminate this dependence,
the ratio $r_\lambda / r_Y$, with $r_Y$ being $r_\lambda$ evaluated at
$\lambda = 1.08~\mu m$ is plotted as a function of $\lambda$ from
0.48 to 1.08 $\mu m$ in Figure 11. The comparison between model and
observed plots then rests on the shape or slope of each curve. Results
from models 1, 6, 9, and 14 are presented to demonstrate the dependence
on $F_K$. In particular, for $T_\ast =$ 4000 K, $r_\lambda / r_Y$ at
0.48 $\mu m$ spans from $8.39$ in model 1 to $1.57$ in model 14.

Looking at the observed plots it is seen that the polynomial fits of
7 stars (BP Tau, CY Tau, DF Tau, DK Tau, DL Tau, DO Tau, LkCa 8), if
replotted with $r_Y$ normalized to 1, will be similar to the model curves.
Two stars (DR Tau, HN Tau) have values of $r_\lambda$, excepting $r_Y$,
that can also be reproduced by models. The $r_\lambda$ plots of two
stars (AA Tau, BM And) are flat from 0.48 to 8.7 $\mu m$. Allowing for
uncertainties in the observed values, they can be considered as being close
to the curve of model 14 whose $F_K$ is the lowest considered. The star
DG Tau has its $r_\lambda$ plot from 0.48 to 0.8 $\mu m$ like a model curve,
but also an upturn in $r_\lambda$ from 0.8 to 0.87 $\mu m$ that
contradicts. The remaining two stars (AS 353A, CW Tau) do not have clearly
defined plots to make a valid comparison, as neighboring $r_\lambda$ values
differ widely.

Overall the accretion shock model can account for a large fraction of the
observed $r_\lambda$ plots. For full affirmation of the model the BJs in
the veiling continua are needed to verify if the $F_K$s implied 
separately by the $r_\lambda / r_Y$
plot and the BJ value are the same. For those objects whose 
$r_\lambda$ values challenge the model their veiling continua need to be
determined from UV to infrared wavelengths in order to compare fully with
those generated by accretion shocks.

\section{Line Fluxes And Shock Parameters}

The accretion shock calculation produces line fluxes in addition to continuum energy distributions. Line emission originating from the post-shock cooling zone have smaller widths in comparison with that originating from the pre-shock photoionization zone which suffer a greater Doppler broadening. They appear as NCs of observed profiles. Among the prominent emergent lines in our model their NC fluxes are also stronger than their broad component (BC) fluxes produced in the pre-shock zone as they are generated at a higher temperature and their optical depths in the post-shock zone are much smaller. The NC fluxes of prominent UV lines, like OVI $\lambda$1034\AA$~$and CIV $\lambda$1549\AA$~$are stronger than those of optical lines, but their observed values depend sensitively on the extinction measure. Optical NC fluxes, on the other hand, are independent of the extinction measure when expressed in equivalent widths relative to the continuum flux densities. The HeI $\lambda$5876\AA$~$NC, in particular, is a prominent line in the model calculation, and among its observed profiles most have either dominant NCs or ones that are easily isolated from BCs (BEK01). It is therefore a prime candidate for verification of the model calculation through comparison with observed data.

Figure 12 shows the dependence on $F_K$ of the $\lambda$5876\AA$~$NC flux (top panel) and its equivalent width relative to the veiling continuum flux density (bottom panel). The NCs of $\lambda$10830\AA$~$and P$\gamma$ are also presented because the $\lambda$10830\AA$~$line, with its lower level being the metastable $2s~{^3S}$ state, bears key information on the physical conditions leading to population of the He excited levels, and P$\gamma$ is always observed along $\lambda$10830\AA$~$. Thus the 5876\AA/10830\AA$~$ and P$\gamma$/10830\AA$~$ flux ratios provide additional checks on the accretion shock model.

All three line fluxes are insensitive to either $u$ or $n_H$ alone, and increase only slowly with increasing $F_K$. An increase in $F_K$ of $\sim$ 40 produces an increase in the HeI $\lambda$5876\AA$~$ NC flux of $\sim$ 3.6. This is because the He and H optical emission occur mostly after the shocked gas has cooled to $\leq$ 2.5$\times 10^4$ K and are close to being thermalized owing to the high electron densities. For a given model flux, the observed flux depends on $\delta$. The equivalent width $W_\lambda ^V$ relative to the veiling continuum flux density, on the other hand, does not. Either result can be used for comparison with observed data.

BEK01 presented in their Figure 4 31 observed NC data in a log[$W_\lambda$x(1 + $r_R$)] versus log $r_R$ plot, where $W_\lambda$ here denotes the NC equivalent width in \AA$~$with respect to the total (photospheric + veiling) continuum, and $r_R = r_\lambda$ at $\lambda$ = 5700 \AA. If $f_\lambda ^\ast$ denotes the photospheric flux density (ergs s$^{-1}$ cm$^{-2}$ \AA$^{-1}$) at wavelength $\lambda$, then the NC flux is given by $W_\lambda$x(1 + $r_\lambda$)$f_\lambda ^\ast$, and the veiling continuum flux density by $r_\lambda f_\lambda ^\ast$ at $\lambda$ = 5876 \AA. Taking $r_R$ as an approximation of $r_{5876\AA}$, the plot then shows the relation between the NC flux and the veiling continuum flux density.

The model results can be presented in a similar way and Figure 13 is a reproduction of the top panel of Figure 4 in BEK01, together with model tracks derived from the model calculations. The latter are obtained with $\delta$ = 0.01, $F_K$ increasing from 4.11x$10^{10}$ (model 14) to 1.66x10$^{12}$ ergs s$^{-1}$ cm$^{-2}$ (model 1) in case a, and for three $T_\ast$ values of 3600, 4000, and 4300 K. The solid line marked $\delta \uparrow$, depicting the trajectory for an equal increment in each co-ordinate, illustrates how each model track will shift for $\delta$s different from 0.01, as both $W_\lambda$x(1 + $r_R$) and $r_R$ are proportional to $\delta$. The area swept by the model tracks as they slide along the direction of varying $\delta$ then outlines the model domain wherein observed data points can be reproduced by one of the 14 models. Alternatively a data point can slide along its own line of $\delta$ variation until it intercepts the model track appropriate to its $T_\ast$ to determine the $F_K$ of the shock model that will reproduce it. The ratio of its $r_R$ to the $r_R$ at the intersection point times 0.01 then yields $\delta$.

The 31 observed stars are sorted into four groups according to the ratio of BC to NC equivalent width (BC/NC). Group 1 has 9 objects, marked +, with each having a prominent NC and no or only a very weak BC (BC/NC $<$ 0.1). Their profiles are shown in Figure 1b of BEK01. Group 4 with 6 objects is characterized by a NC that is absent or much weaker than the BC (BC/NC $>$ 8). Marked $\ast$, they are, from top to bottom, DR Tau, DL Tau, AS 353A, DG Tau, CW Tau, and HN Tau. Each of the remaining objects has easily identifiable NC and BC. They are subdivided into two groups, group 2 with 10 objects having BC/NC between 0.4 and 1.4, and group 3 with 6 objects having BC/NC between 2.5 and 7.2.

Several observations are noted from the spatial distribution of the data points relative to the model tracks: 1) the objects in group 1 all lie within the model domain; 2) 5 of the 6 objects in group 4 are far to the right of the domain as measured by their perpendicular distances from the solid line of varying $\delta$; 3) comparing the four groups, the perpendicular distance of each group from the solid line increases as BC/NC increases.

Two inferences follow from the above findings. First, the group 1 objects or other observed stars with dominant $\lambda$5876\AA$~$NCs can provide a test of the model calculation by verifying if the pair of ($F_K$, $\delta$) values indicated by the NC matches that deduced from modelling the excess continuum. Second, assuming validity of the model calculation, the 14 models clearly fail to replicate some of the data points, particularly those in group 4. One possible explanation is that those points, when extended along their lines of varying $\delta$, just indicate that accretion shocks with $F_K$s much higher than 1.66x10$^{12}$ ergs s$^{-1}$ cm$^{-2}$ and $\delta$s less than 0.01 are needed. This suggestion can be verified readily by examining the BJ and shape of the excess continuum from 3648\AA$~$ to 1 $\mu m$, as both of them correlate strongly with $F_K$ (cf. Figs. 4 \& 11). Another possible explanation is that accretion shocks contribute only a small fraction of the excess continuum in those objects. This fraction can be estimated by extending the data point horizontally to the left, thereby ensuring a constant NC flux, until it intersects the lines of varying $\delta$ extending from the 14 model points that generate the plotted track of the appropriate $T_\ast$. The intersection points provide 14 possible pairs of ($F_K$, $\delta$) that can reproduce the observed NC flux. The ratio of the $r_R$ at each intersection point to the data point $r_R$ then gives the fractional contribution from accretion shocks to the observed excess continuum at 5700\AA$~$ in each case.

The second explanation is preferred here, based on additional information conveyed by the BC profiles, particularly those in groups 4 and 3. Their blue-shifted emission, with blue-shifted centroids and strong blue wings, indicate an origin in outflowing gas (BEK01), and observed  $\lambda$10830\AA$~$ profiles, with broad blue absorptions, provide direct confirmation (EFHK06). Then local excitation calculations find that while temperatures of $\sim 2 \times 10^4$ K are sufficient to produce the observed $\lambda$5876\AA$~$ emission, UV photons that ionize HeI are needed to populate the $2s~^3S$ metastable state via recombination and cascade (Kwan \& Fischer 2011), and such is the case for the NC emission. The absence of or much weaker broad emission observed among group 1 objects indicates that UV photons generated from accretion shocks produce comparatively little emission in either the pre-shock photoionization zone, a result confirmed by our model calculation, or the full magnetospheric accretion columns. Thus the UV photons producing He ionization in the outflow have a different origin. These photons will also cause emission of hydrogen Balmer, Paschen, and other recombination continua and possibly a thermal-like continuum if part of the continuum and line radiation thus produced propagate towards the star and their absorbed fluxes re-emerge as radiation at lower energies. This second origin of UV photons and excess continuum is likely associated with the energy source propelling the outflow.

\section{Discussion}

The main findings of this paper on continuum emission are the BJ - $F_K$ and $r_B - F_K$ 
relations. By means of them the values of $F_K$ and the surface covering
factor $\delta$ can be deduced. Other results are the percentages of
the total veiling flux contributed by the Balmer continuum, spectral lines,
and other constituents. The relative importance between the post-shock 
cooling zone and the pre-shock photoionization zone in generating the
various veiling components, particularly the spectral lines, is also
an outcome.

Comparisons between model and observed veiling continua for 10 T Tauri stars show that our accretion model can reproduce those observed data fairly well. Our derived values of $L_{acc}$ and $\delta$ for BP Tau and the three Orion stars differ considerably from those obtained through application of the shock model of Calvet \& Gullbring (1998). This arises largely from the difference in the dependence of BJ on $F_K$ between the two models and the use of multiple shock components by Ingleby et al. (2014). For the 6 stars in Chamaeleon I our $L_{acc}^{obs}$s are about 27-78\% higher than the corresponding values obtained by Manara et al. (2016).

The use of multiple components to fit the excess continuum is now well established (e.g., Pittman et al. [2022]). This procedure is not without issues. First each component has two parameters, a $F_K$ that governs the shape of the continuum and a $\delta$ that provides the overall magnitude. Each additional component introduces two more free parameters and the higher degrees of freedom in the fitting procedure, while producing a better fit to the data, also reduces the confidence in the derived results. Second, the accuracy of the continuum energy distribution for each $F_K$ calculated by Calvet \& Gullbring (1998) has not been established, so its inaccuracy can by itself cause the poor fit with a single component, further diminishing the merit in the goodness of fit with multiple components. In the end, the derived ($F_K$, $\delta$) values from single or multiple component fitting of the excess continuum need to be confirmed, and spectral line fluxes provide a means. A particular shock model calculation will gain credence only when the set or sets of ($F_K$, $\delta$) obtained from fitting an excess continuum can also reproduce the observed flux of a prominent line formed in accretion shocks.

The alternative shock model calculation presented here is subject to the same test criterion. As noted in $\S$ 5 and seen from Figure 13, objects with dominant HeI $\lambda$5876 \AA$~$ NCs are prime candidates for such a test. The availability of simultaneous continuum and spectral line data through the Penellope and Ullyses programs (Manara et al. 2021, Pittman et al. 2022) makes this test feasible. Model $f_\lambda^V$ and spectral line fluxes for each of the 14 models will be publically available and they can be used to check if the same ($F_K$, $\delta$) set reproduces both the excess continuum and the $\lambda$5876 \AA$~$ NC flux in objects with dominant NCs.
The HeI $\lambda$5876 \AA/HeI $\lambda$10830 \AA$~$ and $P_\gamma$/HeI $\lambda$10830 \AA$~$ flux ratios provide additional tests.

Figure 13 also shows there are objects, particularly those with BCs much stronger than NCs, whose NC flux, given the observed strength of their excess continuum, cannot be reproduced by the 14 models.
They need either accretion shocks with $F_K$s much higher than heretofore calculated or presence of a second source of excess continuum. The first proposition can be tested by verifying if the observed excess continua of those objects have BJs and $r_\lambda$s expected of accretion shocks with very high $F_K$s. 

While a second origin of excess continuum needs to be confirmed, the presence of another source of UV photons besides accretion shocks is clearly indicated on the following grounds. The UV photons produced in the post-shock zone facilitate the optical He emission there by populating the metastable $2s~^3S$ and $2s~^1S$ levels through recombination and cascade that ensue after photoionizations. Collisional excitations from the metastable levels then lead to strong NC emission. Those UV photons that propagate away from the star apparently cause little He emission in either the pre-shock zone or the full magnetospheric accretion columns, as evinced by the 9 objects with dominant NCs and little BCs (BEK01). Among the remaining 22 objects in the data sample, which all have obvious BCs, 11 of them occupy in Figure 13 positions indicating that their NC strengths can be accounted for by the 14 model calculations, just like the 9 objects mentioned earlier. Some of the BCs are stronger than their corresponding NCs. If the accretion shock is the only source of UV photons, it is difficult to fathom how the same realm in ($F_K$, $\delta$) parameter space for the 20 objects generate such drastically different effects at regions farther away. Examination of the 11 BC profiles indicates that several of them have characteristics, like blue-shifted centroids and strong blue wings, that signal line emission from outflowing gas. This kinematic flow is located beyond the accretion columns of infalling gas and UV photons from accretion shocks will have a greater difficulty affecting it because of intervening attenuation and a greater spatial dilution. This predicament faced by accretions shocks as the sole source of UV ionizing photons is even more severe for the remaining 11 objects in the sample, which have BCs more than 2.5 times stronger than NCs and broad blue-shifted profiles indicative of outflow. In contrast, inferring  presence of a second origin of UV ionizing photons whose strength correlates with the flux of the broad He emission makes better sense, and this second origin is likely associated with the energy source propelling the outflow that emits the broad blue-shifted He emission.

The second UV radiation source appears to influence hydrogen line emission also, as the observed P$\gamma$ profiles show outflow characteristics when HeI $\lambda$10830\AA$~$profiles do (cf. Figs. 3 \& 4 of EFHK06). Their fluxes among group 4 objects are also the strongest, with the caveat that the sample is observed much later than the HeI $\lambda$5876\AA$~$sample. Hydrogen Balmer, Paschen, and other recombination continuum emission will be concurrent too. The ratio of P$\gamma$ flux to Balmer continuum flux will likely be different from that generated in the post-shock zone, as the environment around the second UV source is very different. For one thing the UV photons have to affect a larger region, as the HeI $\lambda$5876\AA$~$BC profile indicates that the outflow occupies a fairly large solid angle about the star in order that projections of radial velocities along the line of sight cover a wide span from positive to negative values. Absorption of the ionizing photons also take place at some distance from the star, and the gas densities are likely to be considerably smaller than those at accretion shocks. At the same time line and continuum emission that ensue occur over a bigger region. Those photons propagating towards the stars will be absorbed over a good fraction of the stellar surface. The re-emergent energy flux will likely have an equivalent black-body temperature smaller than that produced in the case of accretion shocks with the same total line and continuum flux incident upon the star. These conjectures can all be verified upon examination of the excess continua observed in stars with HeI $\lambda$5876\AA$~$BCs much stronger than NCs.

The model results reveal that the veiling continuum energy distribution
is insensitive to $u$ or $n_H$ individually. A higher $u$ implies a
higher temperature range in the post-shock zone. Most of the spectral
lines responsible for cooling of the gas at high temperatures deposit
their energies at temperatures $< 3\times 10^4$ K through photoionization
of H and He. The consequent H and He continuum emission from subsequent
radiative recombinations then depend much on the total flux in those 
lines but little on their origins. Free-free emission is produced over
the entire temperature range and, as seen from Figure 1, a greater fraction
of its flux lies at $\lambda < 2000$ \AA$~$ than either the H or thermal
continuum. This fraction depends on $u$, but is difficult to be extracted
as free-free emission constitutes a significant portion of the veiling
continuum only at $\lambda < 1200$ \AA$~$. Unlike continuum
emission, line emission in the post-shock zone occur over a broad range
of density and temperature, owing to the variety of ionic stages and
line excitation parameters. Spectral lines whose fluxes depend differently
on $F_K$ and $u$ can be identified. Their individual fluxes or their
flux ratios will shed light on $u$ besides $F_K$. This finding and other
results deduced through comparisons of model line fluxes with observed
spectral line data are the subject matter of a follow-up paper.

\section{Acknowledgments}

\begin{acknowledgments}

I am grateful to Gopal Narayanan for help on computer equipment, network system and software issues, and a tutorial on Python programming, to Zhiyuan Ji for writing  several Python templates for my plotting needs, and to Bingqing Sun and Yingjie Cheng for help on Python debugging and manuscript preparation and editing on Overleaf. I also thank Suzan Edwards for critical comments on an earlier draft, Will Fischer for information on published literature, and Carlos Manara for providing the relevant data used in Figures 9 and 10. 

\end{acknowledgments}

\section{Data Availability}

The important results of this paper have been presented in Figures and Tables. Extensive results, like the wavelength dependence of the excess continuum of each calculated model, are available upon request. An organized set of all the relevant data will also be available at a website accessed via https://www.stsci.edu/$\sim$wfischer.

\appendix

\section{Atomic Parameters}

The rate coefficients for collisional transitions between HI energy levels are those in Kwan \& Fischer (2011). There it is detailed how they are obtained from Anderson et al. (2002) and Johnson (1972). The rate coefficient for collisional ionization from the ground state is obtained from Arnaud \& Rothenflug (1985). Those for collisional ionizations from excited states are calculated from the formulae given in Johnson (1972). Photoionization cross-sections for the first four levels are gathered from Allen (1973) and radiative recombination coefficients from Seaton (1959).

The HeI atomic parameters used are also those in Kwan \& Fischer (2011). There it is explained how the collision strengths between levels are collected from the Chianti data base (Young et al. 2003) and Sawey \& Berrington (1993), the collisional ionization rate coefficients from Montague, Harrision \& Smith (1984), Taylor, Kingston \& Bell (1979), and Benjamin, Skillman \& Smits (1999), and the radiative recombination coefficients from Benjamin et al. (1999).

For heavy elements their atomic parameters, including level excitation energies, spontaneous emission rates, and rate coefficients for collisional transitions between levels are gathered from Chianti (Young et al. 2003). Their collisional ionization rate coefficients are obtained from Arnaud \& Rothenflug (1985). For hydrogenic ions their radiative recombination rates are calculated from the formula given by Seaton (1959). For the remaining heavy ions their radiative recombination rates are gathered from Arnaud \& Rothenflug (1985), Aldrovandi \& P$\acute{e}$quignot (1973), Shull \& Van Steenberg (1982), and Verner \& Ferland (1996). Dielectronic recombination coefficients are obtained from Romanik (1988) when available, and from Shull \& Van Steenberg (1982) for the others.

Photoionization cross-sections of He and heavy elements used are those in Kwan \& Krolik (1981). There details are given on the use of Seaton (1958) two-term interpolation formula to fit cross-sections as a function of energy, and data collection from Marr \& West (1976), and Reilman \& Manson (1979).

\clearpage

\begin{deluxetable}{cccc}
\tablecaption{Model Parameters}
\tablewidth{0pt}
\tablehead{
\colhead{ Model } &
\colhead{ $n_H$ (cm$^{-3}$) } &
\colhead{ $u$ (km s$^{-1}$) } &
\colhead{ $F_K$ (ergs s$^{-1}$ cm$^{-2}$) }}
\startdata
	\ 1 & 4 $\times$ 10$^{13}$ & 330 & 1.63 $\times$ 10$^{12}$ \\
	\ 2 & 4 $\times$ 10$^{13}$ & 300 & 1.22 $\times$ 10$^{12}$ \\
	\ 3 & 4 $\times$ 10$^{13}$ & 270 & 8.91 $\times$ 10$^{11}$ \\
	\ 4 & 4 $\times$ 10$^{13}$ & 240 & 6.26 $\times$ 10$^{11}$ \\
	\ 5 & 4 $\times$ 10$^{13}$ & 200 & 3.62 $\times$ 10$^{11}$ \\
	\ 6 & 1 $\times$ 10$^{13}$ & 330 & 4.07 $\times$ 10$^{11}$ \\
	\ 7 & 1 $\times$ 10$^{13}$ & 300 & 3.06 $\times$ 10$^{11}$ \\
	\ 8 & 1 $\times$ 10$^{13}$ & 270 & 2.23 $\times$ 10$^{11}$ \\
	\ 9 & 1 $\times$ 10$^{13}$ & 240 & 1.56 $\times$ 10$^{11}$ \\
	\ 10 & 1 $\times$ 10$^{13}$ & 200 & 9.05 $\times$ 10$^{10}$ \\
	\ 11 & 2.5 $\times$ 10$^{12}$ & 330 & 1.02 $\times$ 10$^{11}$ \\
	\ 12 & 2.5 $\times$ 10$^{12}$ & 300 & 7.64 $\times$ 10$^{10}$ \\
	\ 13 & 2.5 $\times$ 10$^{12}$ & 270 & 5.57 $\times$ 10$^{10}$ \\
	\ 14 & 2.5 $\times$ 10$^{12}$ & 240 & 3.91 $\times$ 10$^{10}$ \\
\enddata
\end{deluxetable}

\clearpage

\begin{sidewaystable*}
	\caption{Selected Model Results}
	\begin{tabular} {cccccc|ccccc}
	\toprule
	\multicolumn{6}{c}{Case a}&\multicolumn{5}{c}{Case b}\\
	\hline
	\colhead{Model}& \colhead{$F_V$}& \colhead{$F_{thrm}\over{F_V}$}&
		\colhead{$F_{Bal}\over{F_V}$}& 
		\colhead{$F_{slns}\over{F_V}$}& \colhead{BJ}&
		\colhead{$F_V$}& \colhead{$F_{thrm}\over{F_V}$}&
		\colhead{$F_{Bal}\over{F_V}$}&
		\colhead{$F_{slns}\over{F_V}$}& \colhead{BJ}\\
	  &(ergs s$^{-1}$ cm$^{-2}$)&  &  &  &  &
		(ergs s$^{-1}$ cm$^{-2}$)&  &  &  &  \\
	\hline
		1& 1.66$\times10^{12}$& 0.466& 0.156& 0.108& 1.384&
		1.557$\times10^{12}$& 0.46& 0.167& 0.092& 1.456\\
		2& 1.26$\times10^{12}$& 0.475& 0.151& 0.127& 1.512&
		1.194$\times10^{12}$& 0.466& 0.163& 0.106& 1.569\\
		3& 9.269$\times10^{11}$& 0.48& 0.144& 0.152& 1.627&
		8.904$\times10^{11}$& 0.471& 0.158& 0.124& 1.698\\
		4& 6.596$\times10^{11}$& 0.482& 0.135& 0.183& 1.741&
		6.405$\times10^{11}$& 0.472& 0.154& 0.145& 1.85\\
		5& 3.907$\times10^{11}$& 0.481& 0.115& 0.243& 1.891&
		3.813$\times10^{11}$& 0.469& 0.144& 0.184& 2.108\\
		6& 4.159$\times10^{11}$& 0.492& 0.139& 0.142& 1.736&
		3.956$\times10^{11}$& 0.481& 0.153& 0.123& 1.835\\
		7& 3.147$\times10^{11}$& 0.493& 0.135& 0.162& 1.927&
		3.022$\times10^{11}$& 0.481& 0.15& 0.139& 2.051\\
		8& 2.316$\times10^{11}$& 0.492& 0.129& 0.188& 2.164&
		2.243$\times10^{11}$& 0.48& 0.146& 0.157& 2.338\\
		9& 1.646$\times10^{11}$& 0.489& 0.119& 0.222& 2.464&
		1.606$\times10^{11}$& 0.477& 0.142& 0.178& 2.755\\
		10& 9.758$\times10^{10}$& 0.483& 0.099& 0.284& 3.021&
		9.545$\times10^{10}$& 0.467& 0.132& 0.223& 3.693\\
		11& 1.038$\times10^{11}$& 0.502& 0.124& 0.179& 2.859&
		9.969$\times10^{10}$& 0.489& 0.139& 0.16& 3.128\\
		12& 7.858$\times10^{10}$& 0.499& 0.12& 0.2& 3.456&
		7.592$\times10^{10}$& 0.486& 0.136& 0.177& 3.817\\
		13& 5.778$\times10^{10}$& 0.494& 0.114& 0.229& 4.334&
		5.616$\times10^{10}$& 0.482& 0.133& 0.197& 4.872\\
		14& 4.109$\times10^{10}$& 0.489& 0.104& 0.264& 5.634&
		4.017$\times10^{10}$& 0.476& 0.128& 0.222& 6.553\\
	\hline
	\end{tabular}
\end{sidewaystable*}

\clearpage

\begin{sidewaystable*}
	\caption{Model Results With Inclusion Of Emergent Flux From 
	Stellar Interior}
        \begin{tabular}{cccc|ccc|cccc|cccc}
	\toprule
		\multicolumn{4}{c}{BJ}&\multicolumn{3}{c}{PJ}&\multicolumn
		{4}{c}{$r_B$}&\multicolumn{4}{c}{$r_R$}\\
	\hline
	\colhead{Model}&
	\colhead{C1}&
	\colhead{C$2^{\prime}$}&
	\colhead{C$1^{\prime}$}&
	\colhead{C1}&
	\colhead{C$2^{\prime}$}&
	\colhead{C$1^{\prime}$}&
	\colhead{C1}&
	\colhead{C$1^{\prime}$}&
	\colhead{C2}&
	\colhead{C$2^{\prime}$}&
	\colhead{C1}&
	\colhead{C$1^{\prime}$}&
	\colhead{C2}&
	\colhead{C$2^{\prime}$}\\
	\hline
		9 & 2.464 & 2.242 & 2.144 & 1.144 & 1.133 & 1.127 &
		0.194 & 0.232 & 0.473 & 0.533 & 0.109 & 0.126 & 0.221 & 0.242\\
		12 & 3.456 & 2.711 & 2.446 & 1.143 & 1.116 & 1.107 &
		0.072 & 0.108 & 0.175 & 0.231 & 0.049 & 0.0673 & 0.099 &
		0.123\\
		14 & 5.634 & 3.343 & 2.755 & 1.137 & 1.093 & 1.079 &
		0.0239 & 0.053 & 0.0582 & 0.103 & 0.02 & 0.0381 & 0.0405 & 
		0.064\\
        \hline
	\end{tabular}
\end{sidewaystable*}

\clearpage

\begin{sidewaystable*}
	\caption{Comparison Between Observed And Model Results}
	\begin{tabular} {cccccccc|cccccc}
        \toprule
	\multicolumn{8}{c}{Data}&\multicolumn{6}{c}{Model}\\
	\hline
        \colhead{star}&
	\colhead{$d$}&
        \colhead{$R_\ast$}&
	\colhead{$T_\ast$}&
	\colhead{$r_B$}&
	\colhead{$\epsilon_{bb}$}&
	\colhead{$L_{acc}^{obs}$}&
	\colhead{$L_{acc}$}&
	\colhead{No.}&
	\colhead{BJ}&
	\colhead{$F_K$}&
	\colhead{$\delta_{bb}$}&
	\colhead{$L_{acc}^{obs}$}&
	\colhead{$L_{acc}$}\\
	  &(pc) &($R_\odot$) &($10^3$K) &  &  &($L_\odot$) &(L$_\odot$) &  &  &
		(erg s$^{-1}$ cm$^{-2}$) &  &($L_\odot$) &($L_\odot$)\\
        \hline
		BP Tau & 140 & 1.99 & 4.0 & 1 & 0.42 &  &
		0.14 & 3 & 1.63 & 9.27$\times10^{11}$ & 0.0077&
		0.185 & 0.44\\
		SO 1036 & 440 & 2.9 & 3.7 & 0.9 & 0.4 &  & 0.1 & 9 & 2.46 &
		1.65$\times10^{11}$ & 0.0242 & 0.212 & 0.53\\
		SO 540 & 440 & 2.0 & 4.0 & 0.25 & 0.48 &  & 0.1 & 7 & 1.93 &
		3.15$\times10^{11}$ & 0.0057 & 0.054 & 0.112\\
		CVSO 58 & 400 & 1.5 & 4.0 & 1.9 & 0.6 &  & 0.3 & 1 & 1.4 &
		1.66$\times10^{12}$ & 0.0086 & 0.3 & 0.5\\
		TW Cha& 160& 1.25& 4.06& 0.22& 0.71& 0.022&  &
		8 & 2.16 & 2.32$\times10^{11}$ & 0.0082 &
		0.033 & 0.047\\
		T3 & 160 & 0.86 & 4.06& 1.25& 0.7& 0.056&  &
		7 & 1.93 & 3.15$\times10^{11}$ & 0.032 &
		0.083 & 0.118\\
		VW Cha& 160& 2.6& 4.06& 0.36& 0.67& 0.166&  & 2& 1.51&
		1.26$\times10^{12}$& 0.00233& 0.21& 0.31\\
		ESOH$\alpha$562& 160& 0.84& 3.705& 0.55& 0.63& 0.0098&  & 8&
		2.16& 2.32$\times10^{11}$& 0.0096& 0.0157& 0.025\\
		T3-B& 160& 1.25& 3.415& 1.65& 0.42& 0.022&  & 6& 1.74&
		4.16$\times10^{11}$& 0.0069& 0.029& 0.07\\
		T40& 160& 1.74& 3.78& 3.2& 0.62& 0.33&  & 1& 1.4&
		1.66$\times10^{12}$& 0.0091& 0.44& 0.71\\

	\hline
	\end{tabular}
\end{sidewaystable*}

\clearpage
\begin{figure}
\plotone{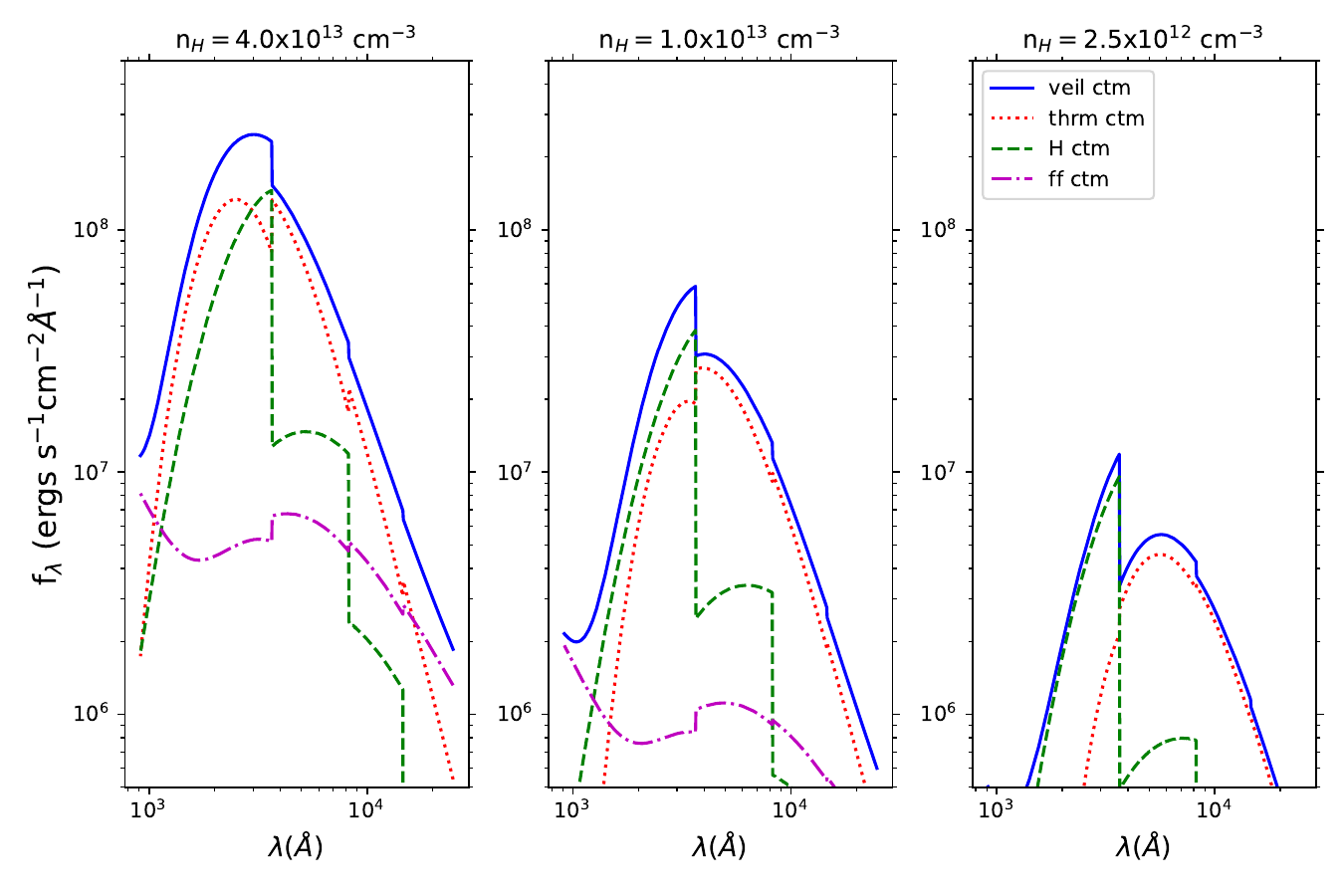}
	\caption{Emergent continuum energy distributions of models 2, 7, and
	12, with $u = 300$ km s$^{-1}$ and three $n_H$ values in case a. 
	Their constituents of thermal continua, hydrogen recombination
	continua comprising Balmer, Paschen, and Brackett components, and
	free-free continua are also presented.}
\end{figure}

\clearpage
\begin{figure}
\includegraphics[width=0.95\linewidth]{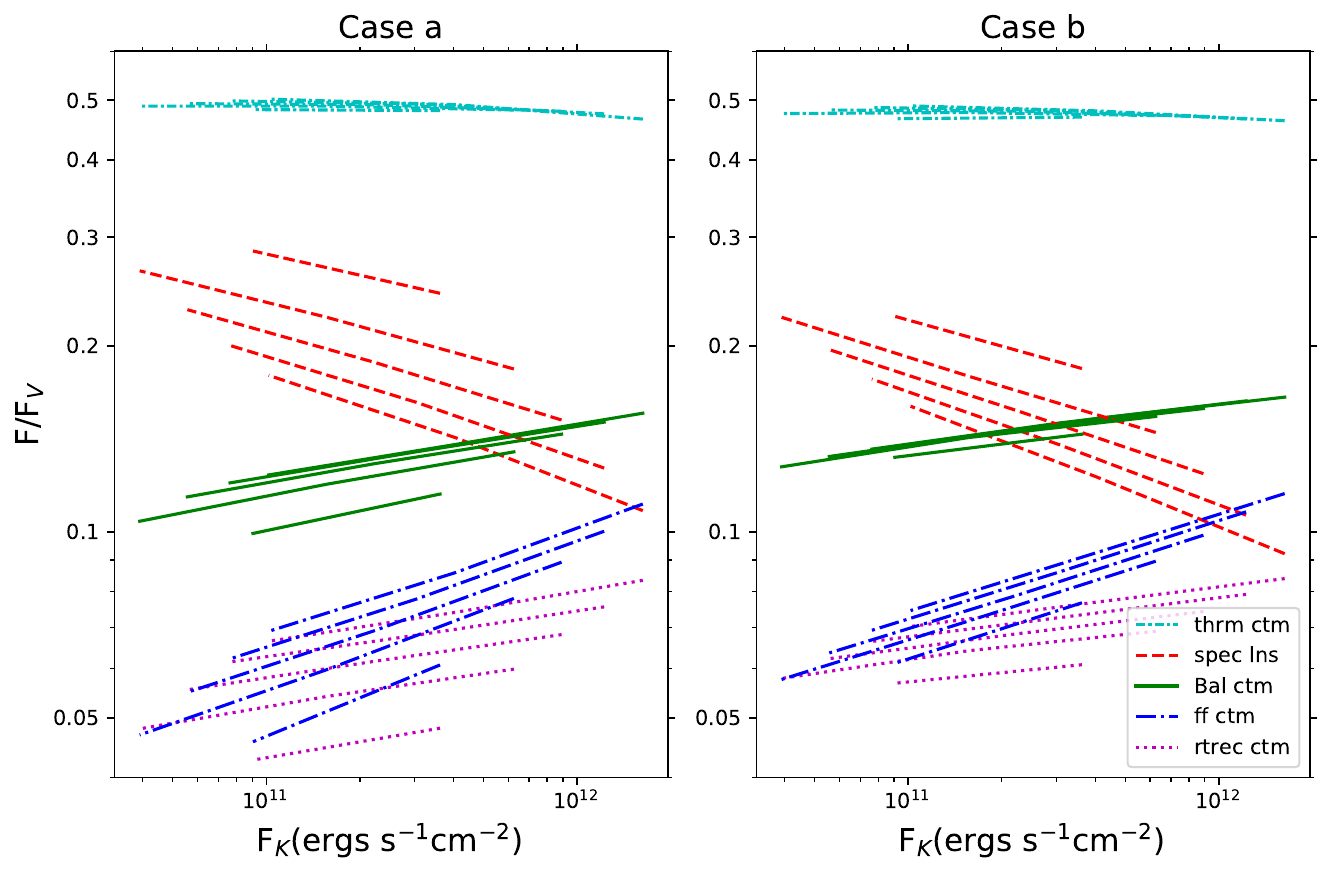}
	\caption{Fraction of the total veiling flux contributed by its
	individual constituents as a function of $F_K$. Each of the 5 lines
	for each constituent connects the data points at a fixed model
	parameter of $u$. From top to bottom the set of 5 lines indicates the fraction contributed by the thermal continuum, spectral lines, Balmer continuum, free-free continuum, and the summed continuum of radiative recombinations of H to levels n$\geq$5, HeI to n$\geq$2, and HeII to n$\geq$3. Cases a and b refer to calculations with complete and no suppression of dielectronic recombination respectively.}
\end{figure}

\clearpage
\begin{figure}
\centering
\includegraphics[width=0.95\linewidth]{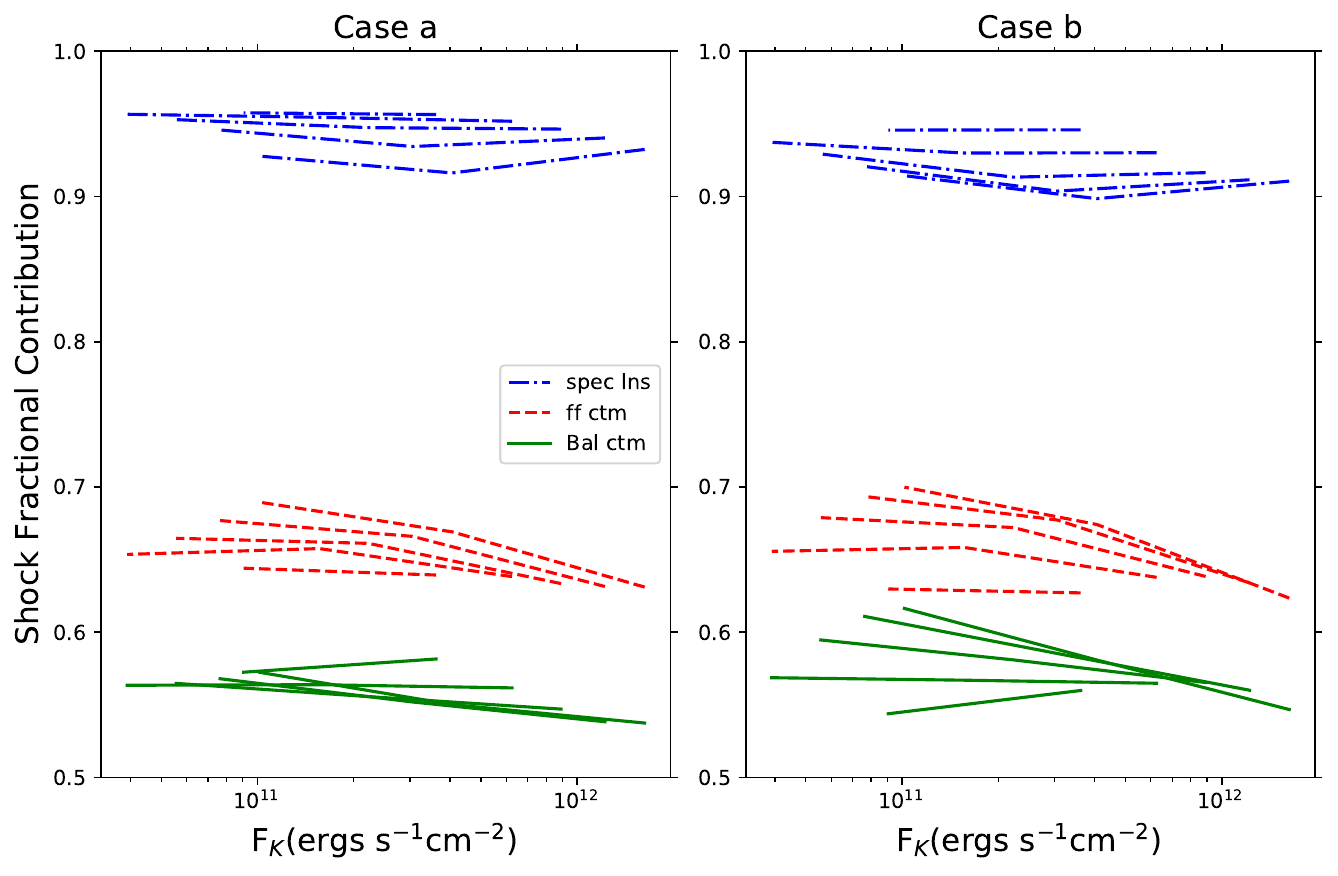}
	\caption{Fraction of the emergent flux of a particular component
	contributed by the post-shock cooling zone as a function of
	$F_K$. Same annotation for the 5 lines as in
	Fig. 2. From top to bottom the set of 5 lines refers to the component of spectral lines, free-free continuum, and Balmer continuum.}
\end{figure}

\clearpage
\begin{figure}
\centering
\includegraphics[width=0.95\linewidth]{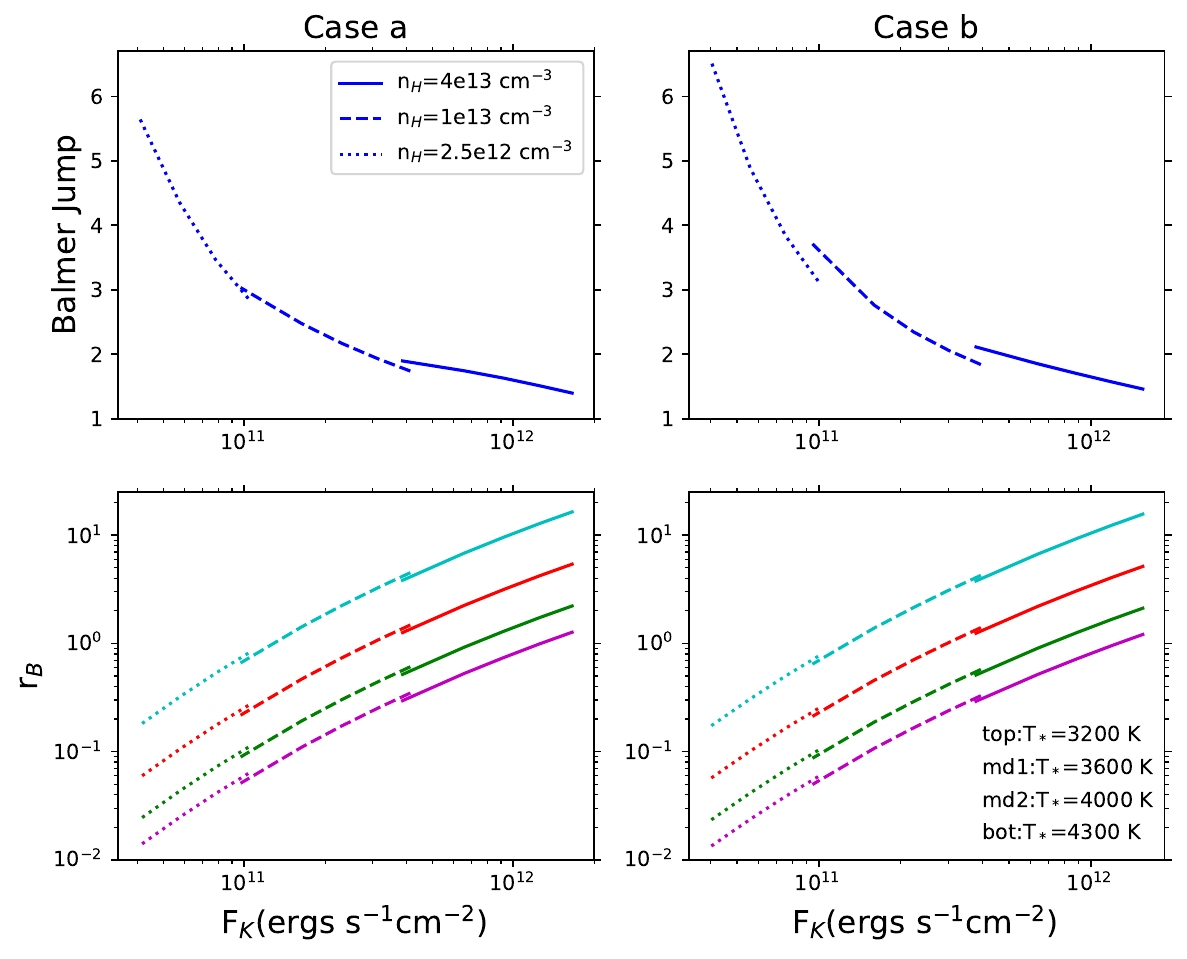}
	\caption{Dependence of the Balmer jump and $r_B$ on $F_K$.}
\end{figure}

\clearpage
\begin{figure}
\centering
\includegraphics[width=0.85\linewidth]{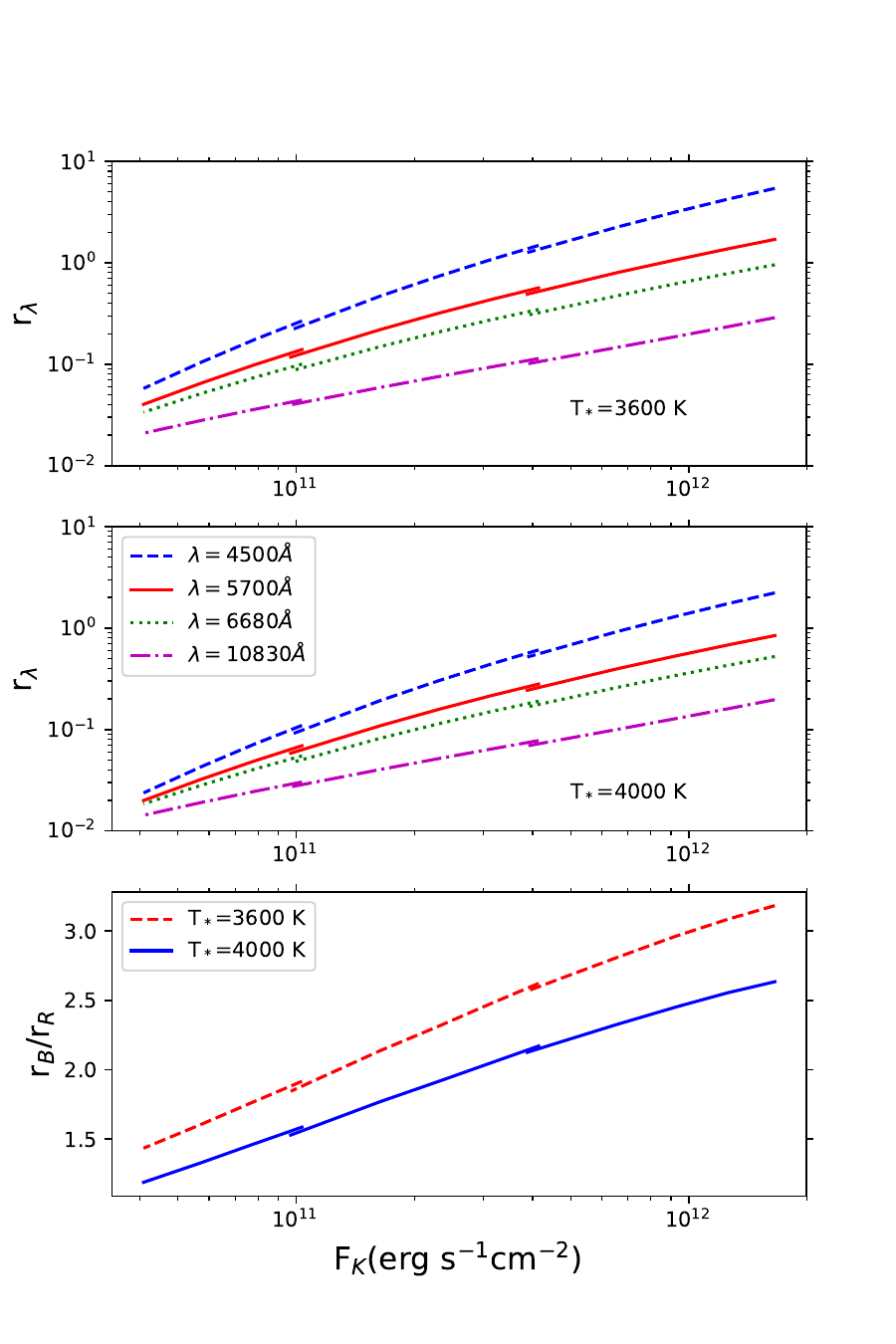}
	\caption{Dependence of $r_\lambda$ on $F_K$ for 4 $\lambda$s
	in case a, assuming $T_\ast =$ 3600 K (top panel) and 4000 K
	(middle panel). Corresponding dependence of $r_B / r_R$ on
	$F_K$ for the 2 $T_\ast$s (bottom panel).}
\end{figure}

\clearpage
\begin{figure}
\centering
\includegraphics[width=0.75\linewidth]{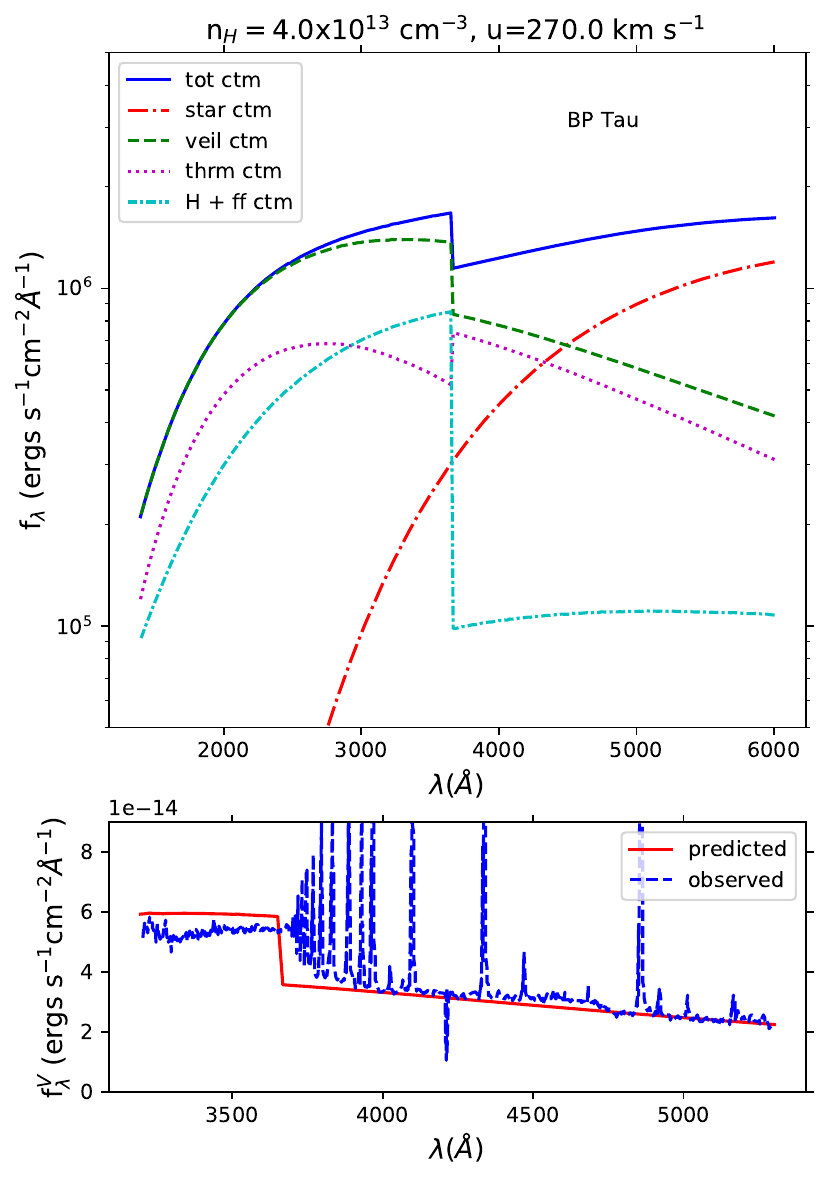}
	\caption{Total continuum spectrum from shock model
	for BP Tau. Top panel shows a 4000 K
	black-body spectrum,  
	the emergent continuum spectrum
	of model 3 multiplied by $\delta_{bb}$ to simulate
	the veiling continuum spectrum, and their sum to produce
	the model total continuum spectrum. Also shown are
	the veiling continuum constituents, namely, the thermal
	continuum and the summed free-free, hydrogen Balmer,
	Paschen and Brackett continua. Bottom panel shows the predicted 
	veiling continuum based
	on adopted values of $R_\ast$, $d$ and $\epsilon_{bb}$ (cf. Table 4), and the observed excess continuum reproduced from Fig. 7 of Gullbring et al. (1998).}
\end{figure}

\clearpage
\begin{figure}
\centering
\includegraphics[width=0.85\linewidth]{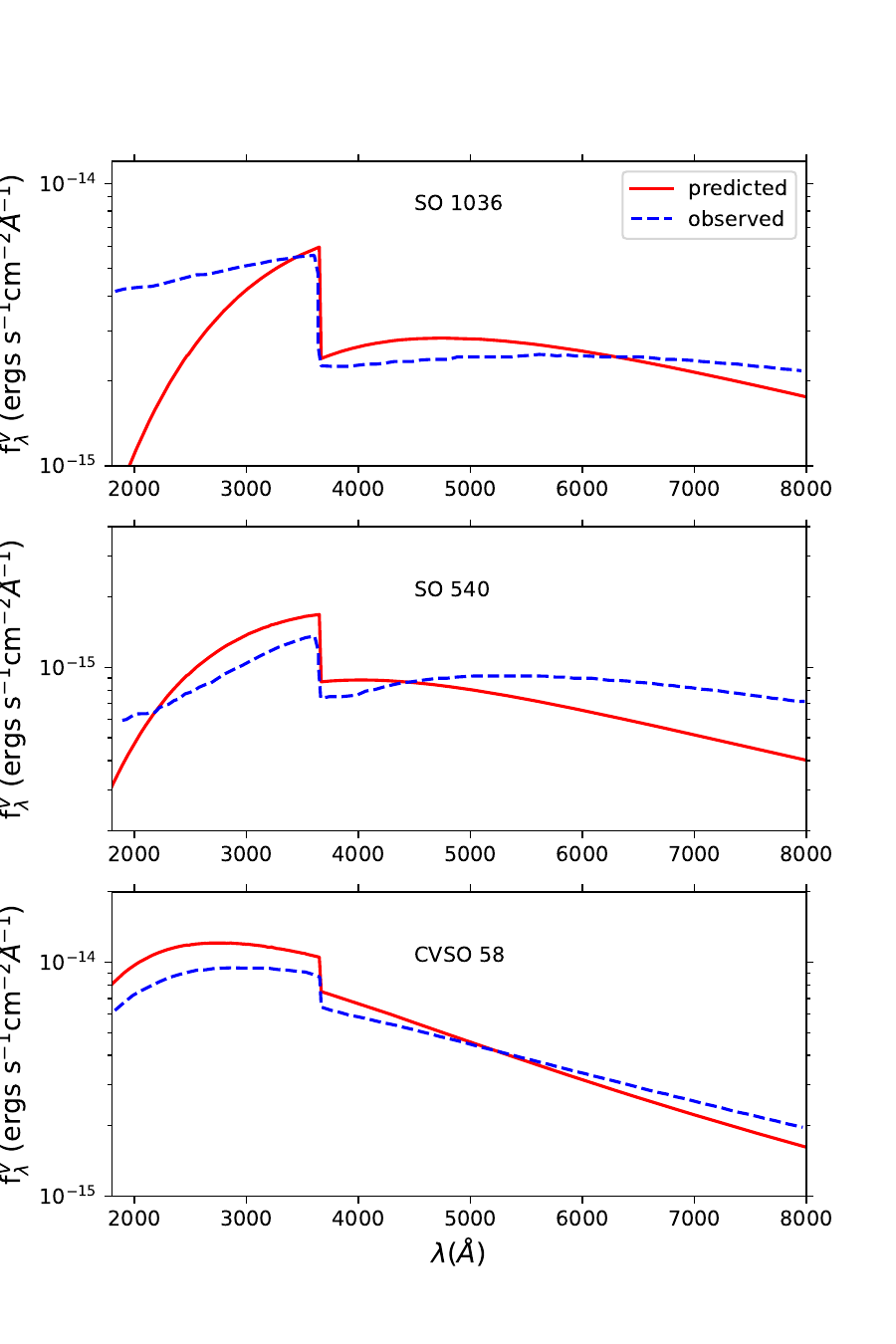}
\caption{Predicted excess continuum energy distributions
	from shock models (solid lines) simulating the excess continua (dashed lines) of 3 Orion stars.
    See Table 4 for relevant
	model parameters and adopted stellar properties. Observed data reproduced from Fig. 5 of Ingleby et al. (2014).}
\end{figure}

\clearpage

\begin{figure}
\centering
\includegraphics[width=0.85\linewidth]{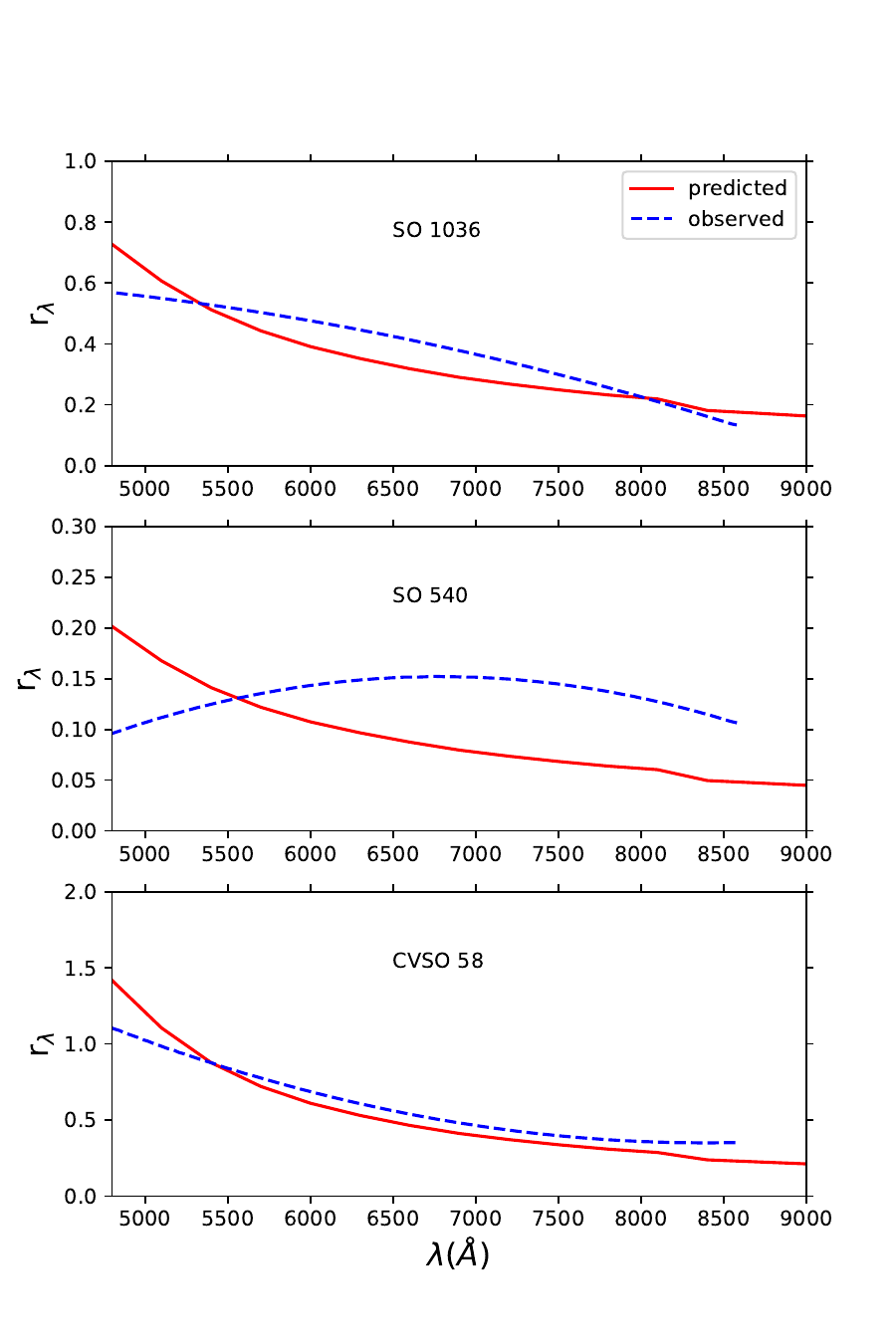}
\caption{Predicted dependence of $r_\lambda$ on $\lambda$ (solid   lines) for the
	same 3 stars in Fig. 7. Also shown are fits to observed data points (dashed lines) reproduced from Fig. 3 of Ingleby et al. (2014).}
\end{figure}

\clearpage

\begin{figure}
\centering
\includegraphics[width=0.85\linewidth]{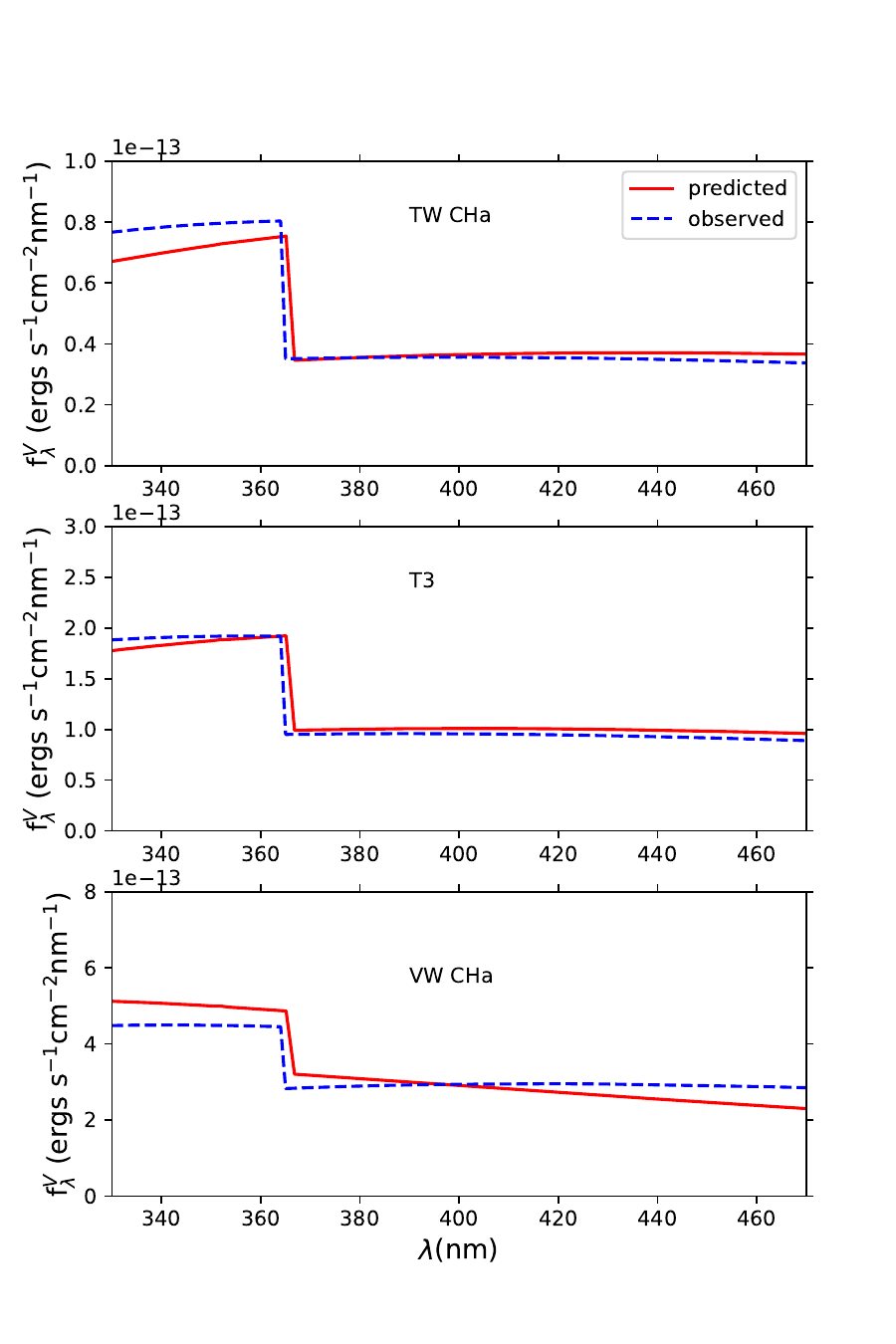}
	\caption{Predicted excess continuum energy distributions from 
	shock models (solid lines) simulating the excess continua (dashed lines) of 3 stars
	in data sample of Manara et al. (2016). See Table 4 for relevant
	model parameters and adopted stellar properties. Observed data kindly provided by Carlos Manara.}
\end{figure}

\clearpage
\begin{figure}
\centering
\includegraphics[width=0.85\linewidth]{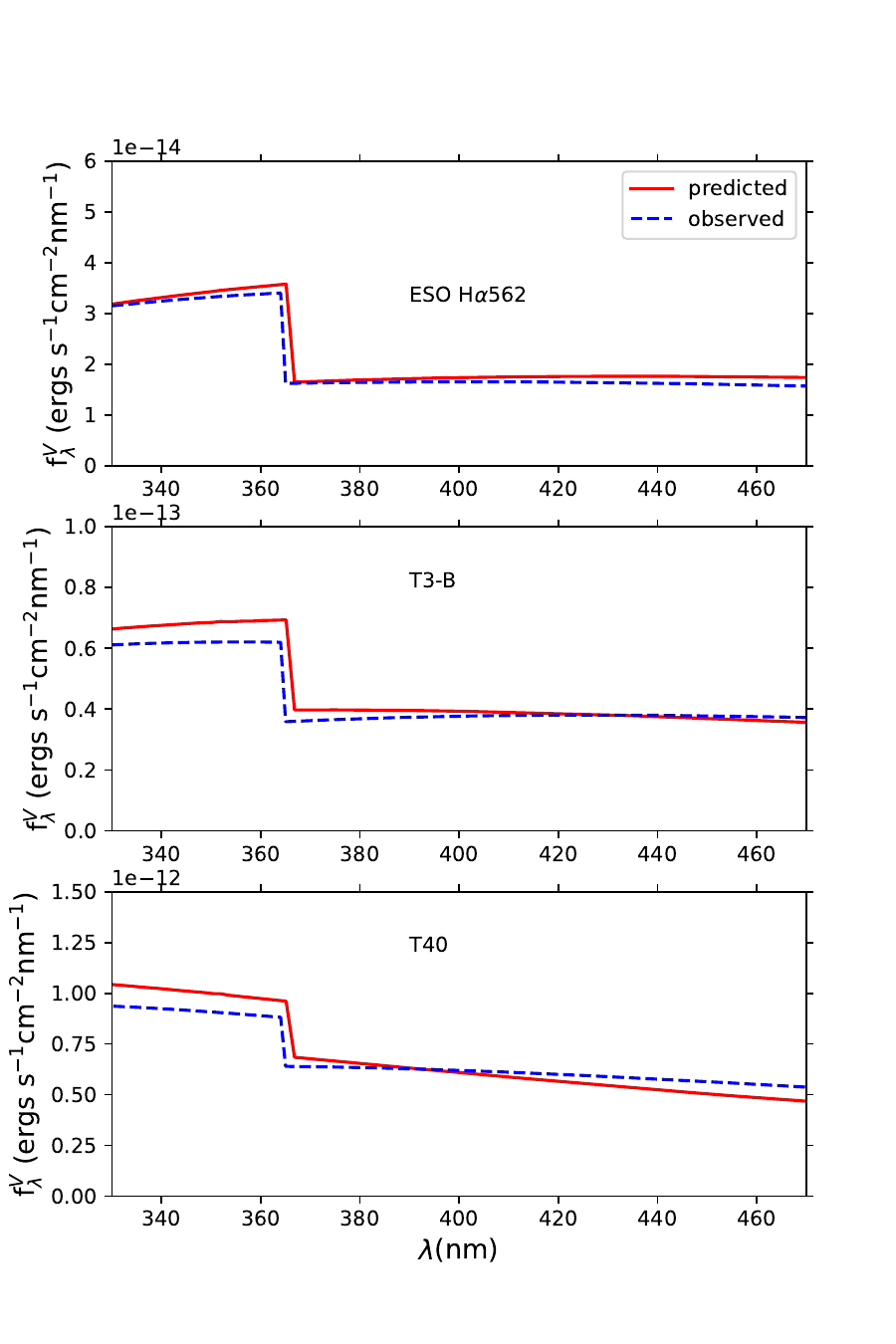}
	\caption{Same as Fig. 9, for another 3 stars in data sample
	of Manara et al. (2016).}
\end{figure}

\clearpage
\begin{figure}
\centering
\includegraphics[width=0.85\linewidth]{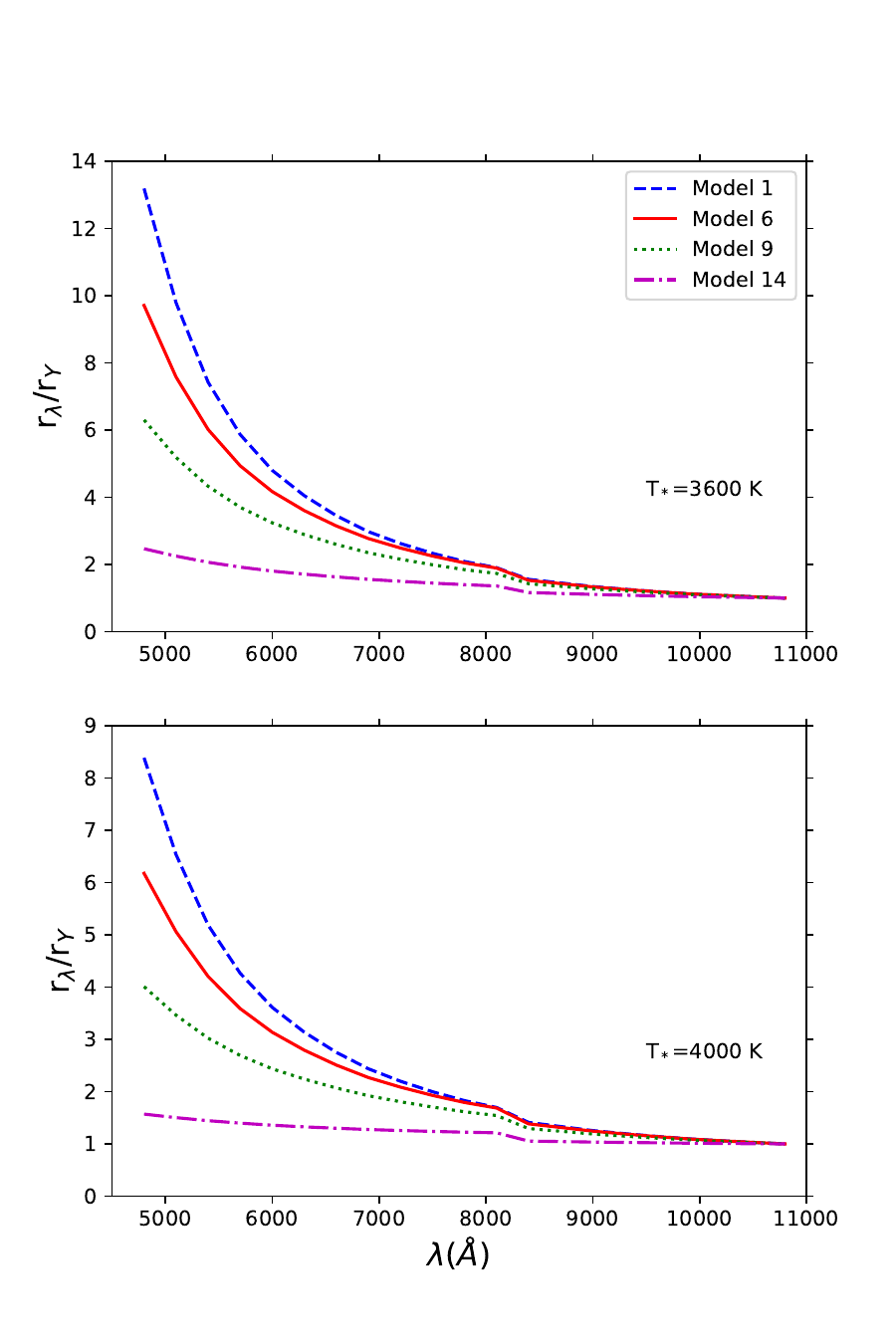}
	\caption{Dependence of $r_\lambda / r_Y$ on $\lambda$ in 4 shock
	models.}
\end{figure}

\clearpage
\begin{figure}
\centering
\includegraphics[width=0.85\linewidth]{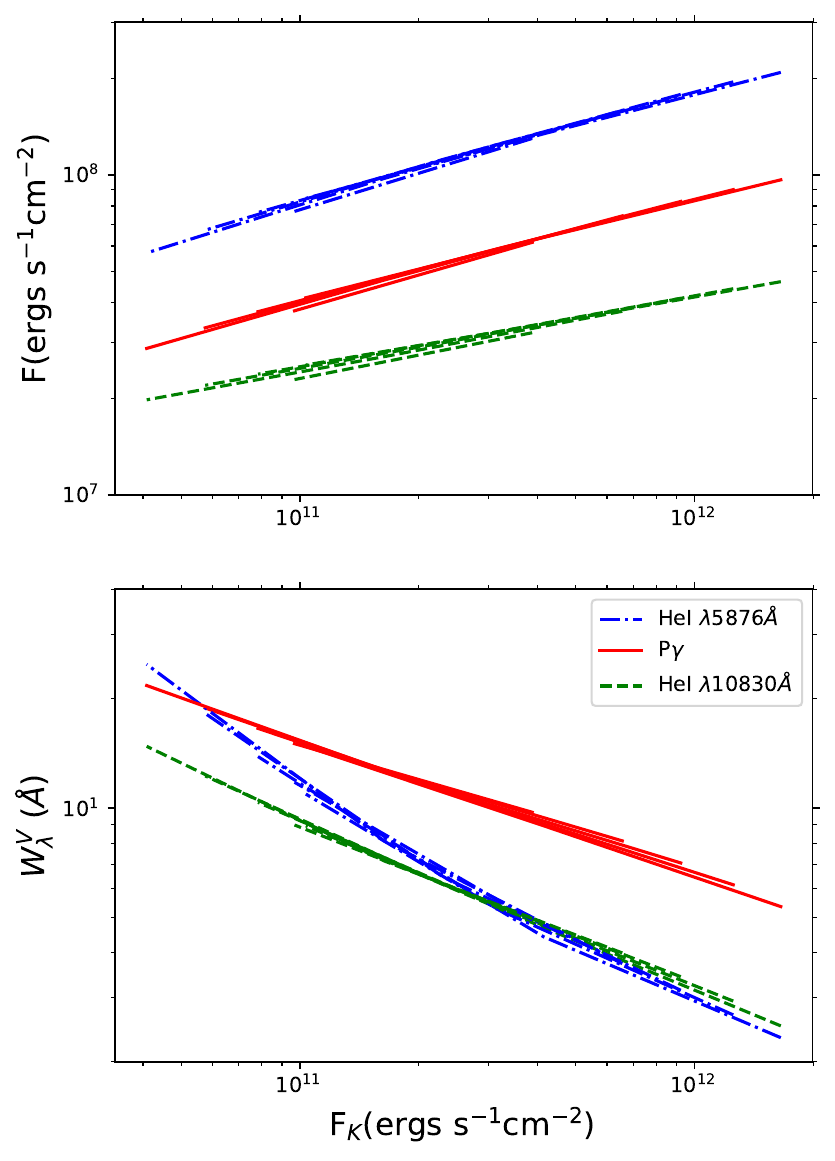}
\caption{Dependence of line flux (top panel) and equivalent width relative to veiling continuum (bottom panel) on $F_K$. Each spectral line plot consists of 5 lines showing results for the 5 model values of $u$.}
\end{figure}

\clearpage
\begin{figure}
\centering
\includegraphics[width=0.95
\linewidth]{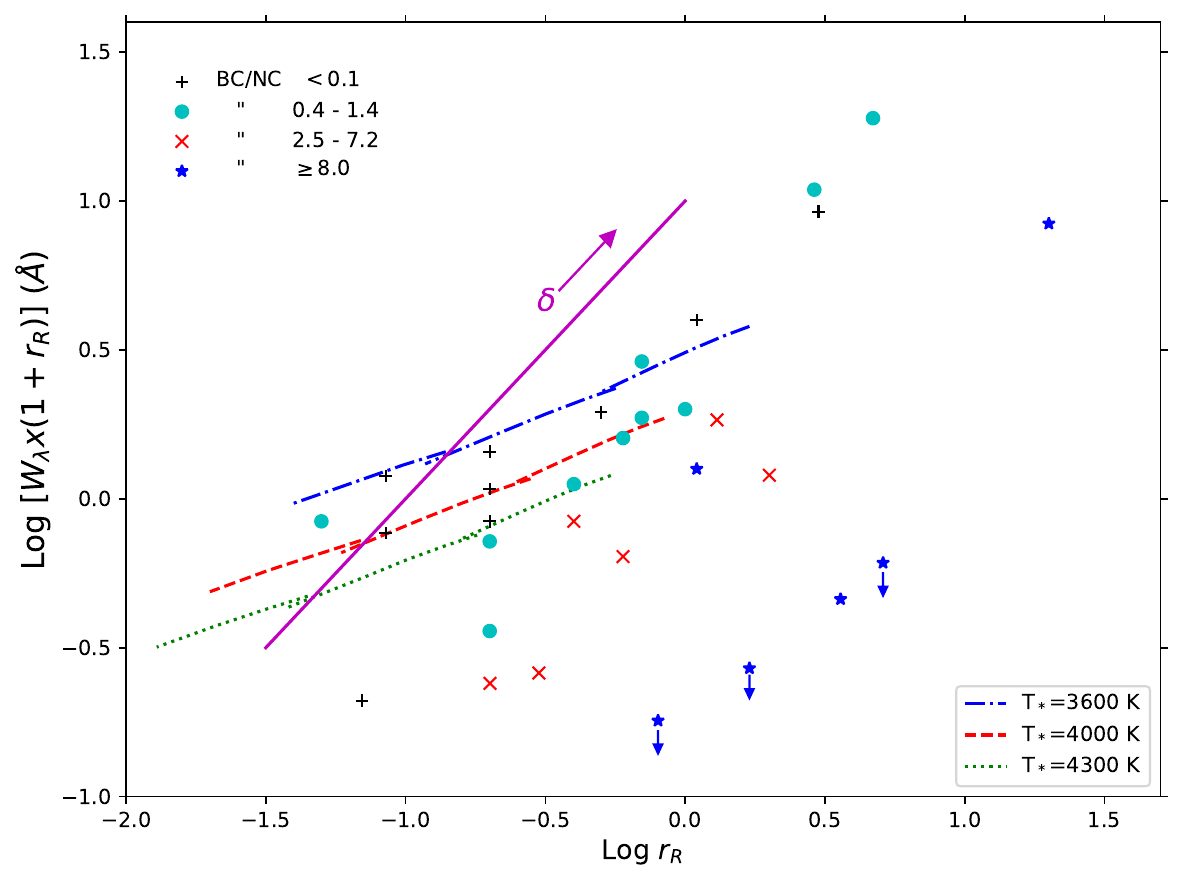}
    \caption{Plot of HeI $\lambda$5876\AA$~$ NC equivalent width $W_\lambda$ times (1 + $r_R$) versus $r_R$, with data points from BEK01 and model tracks derived from accretion shock calculations with $\delta = 0.01$, $F_K$ increasing from the model 14 value to the model 1 value in case a, and 3 $T_\ast$s. The solid line illustrates how the model tracks will move along for different $\delta$s, as both $r_R$ and $W_\lambda$ x (1 + $r_R$) are $\propto \delta$. Data points are marked by different symbols according to the ratio of observed BC to NC equivalent width.}
\end{figure}


\begin{references}
\reference{}
Adams T. F., 1972, ApJ, 174,439

\reference{}
Aldrovandi S. M. V., P$\acute{e}$quignot D., 1973, A\&A, 25, 137

\reference{}
Alencar S. H. P., Basri G., 2000, AJ, 119,1881

\reference{}
Allen C. W., 1973, Astrophysical Quantities, Athlone, London

\reference{}
Anderson H., Ballance C. P., Badnell N. R., Summers H. P., 2002, J. Phys. B, 35, 1613

\reference{}
Ardila D. R., Herczeg G. J., Gregory S. G., et al., 2013, ApJS, 207,1

\reference{}
Arnaud M., Rothenflug R., 1985, A\&AS, 60, 425

\reference{}
Avery L. H., House L. L., 1968, ApJ, 152, 493

\reference{}
Basri G., Batalha C., 1990, ApJ, 363, 654

\reference{}
Benjamin R. A., Skillman E. D., Smits D. P., 1999, ApJ, 514,307

\reference{}
Beristain G., Edwards S., Kwan J., 2001, ApJ, 551,1037 (BEK01)

\reference{}
Calvet N., Gullbring E., 1998, ApJ, 509, 802

\reference{}
Davidson K., 1975, ApJ, 195, 285

\reference{}
Dodin A., 2018, MNRAS, 475, 4367

\reference{}
Edwards S., Fischer W., Hillenbrand L., Kwan J., 2006, ApJ, 646, 319 (EFHK06)

\reference{}
Fischer W., Edwards S., Hillenbrand L., Kwan J., 2011, ApJ, 730, 73

\reference{}
Gullbring E., Calvet N., Muzerolle J., Hartmann L., 2000, ApJ, 544, 927

\reference{}
Gullbring E., Hartmann L., Brice$\tilde{n}$o C., Calvet N., 1998, ApJ, 492,323

\reference{}
Hartmann L., Herczeg G. J., Calvet N., 2016, ARA\&A, 54, 135

\reference{}
Ingleby L., Calvet N., Herczeg G. J., et al., 2013, ApJ, 767, 112

\reference{}
Ingleby L., Calvet N., Hern$\acute{a}$ndez J., et al., 2014, ApJ, 790, 47

\reference{}
Johns-Krull C. M., Chen W., Valenti, J. A., et al., 2013, ApJ, 765, 11

\reference{}
Johnson L. C., 1972, ApJ, 174, 227

\reference{}
Kwan J., Fischer W., 2011, MNRAS, 411,2383

\reference{}
Kwan J., Krolik J. H., 1981, ApJ, 250, 478

\reference{}
Lodders K., 2020, Solar Elemental Abundances, in Oxford Research Encyclopedia of Planetary Science

\reference{}
Manara C. F., Fedele D., Herczeg G. J., Teixeira P. S., 2016, A\&A, 585, A136

\reference{}
Manara C. F., Frasca A., Venuti L., et al., 2021, A\&A, 650, A196

\reference{}
Marr G. V., West J. B., 1976, Atomic Data and Nuclear Data Tables, 18, 497

\reference{}
Montague R. G., Harrison M. F. A., Smith A. C. H., 1984, J Phys. B., 17, 3295

\reference{}
Pittman C. V., Espaillat C. C., Robinson C. E., et al., 2022, AJ, 164, 201

\reference{}
Raymond J. C., Smith B. S., 1977, ApJS, 35, 419

\reference{}
Reilman R. F., Manson S. T., 1979, ApJS, 40, 815

\reference{}
Robinson C. E., Espaillat C. C., 2019, ApJ, 874, 129

\reference{}
Romanik C. J., 1988, ApJ, 330, 1022

\reference{}
Sawey P. M. J., Berrington K. A., 1993, At. Data Nucl. Data Tables, 55, 81

\reference{}
Seaton M. J., 1958, Rev. Mod. Phys., 30, 979

\reference{}
Seaton M. J., 1959, MNRAS, 119, 81

\reference{}
Shull J. M., Van Steenberg M., 1982, ApJS, 48, 95

\reference{}
Sutherland R. S., 1998, MNRAS, 300, 321

\reference{}
Taylor I. R., Kingston A. E., Bell K. L., 1979, J. Phys. B, 12, 3093

\reference{}
Verner D. A., Ferland G. J., 1996, ApJS, 103,467

\reference{}
Young P. R., Zanna G. D., Landi E., Dere K. P., Mason H. E., Landini M., 2003, ApJS, 144, 135

\end{references}
\end{document}